\documentclass[showpacs,twocolumn,pra,superscriptaddress,notitlepage]{revtex4-2}
\usepackage{graphicx}
\usepackage{physics}
\usepackage{wrapfig}
\usepackage{changepage}
\usepackage{mathtools}
\usepackage{amsmath,amssymb,amsthm,mathrsfs,amsfonts,dsfont,mathtools}
\usepackage[caption=false]{subfig}
\usepackage{epsfig}
\usepackage{color,xcolor}
\usepackage{adjustbox}
\usepackage{tabularx}
\usepackage{array}
\usepackage{multirow}
\usepackage{tabu}
\usepackage{bm}
\usepackage{enumerate}
\usepackage{hyperref}
\usepackage{newtxtext,newtxmath}
\usepackage{makecell}
\usepackage{url}
\usepackage{booktabs}
\usepackage{nicefrac}
\usepackage{comment}

\newcommand{\PP}[1][]{
  \ifthenelse{\isempty{#1}}
    {\mathbbm{P}}
    {\mathbbm{P}\left[#1\right]}
}
\newcommand{\EE}[1][]{
  \ifthenelse{\isempty{#1}}
    {\mathbbm{E}}
    {\mathbbm{E}\left[#1\right]}
}

\DeclareMathOperator*{\argmin}{arg\,min}
\usepackage{bbm}

\begin{document}

\title{
   Neural Quantum Embedding:
   Pushing the Limits of Quantum Supervised Learning
}

\author{Tak Hur}
\affiliation{Department of Statistics and Data Science, Yonsei University, Seoul 03722, Republic of Korea}
\author{Israel F. Araujo}
\affiliation{Department of Statistics and Data Science, Yonsei University, Seoul 03722, Republic of Korea}
\author{Daniel K. Park}
\email{dkd.park@yonsei.ac.kr}
\affiliation{Department of Statistics and Data Science, Yonsei University, Seoul 03722, Republic of Korea}
\affiliation{Department of Applied Statistics, Yonsei University, Seoul 03722, Republic of Korea}

\begin{abstract}
Quantum embedding is a fundamental prerequisite for applying quantum machine learning techniques to classical data, and has substantial impacts on performance outcomes. In this study, we present Neural Quantum Embedding (NQE), a method that efficiently optimizes quantum embedding beyond the limitations of positive and trace-preserving maps by leveraging classical deep learning techniques. NQE enhances the lower bound of the empirical risk, leading to substantial improvements in classification performance. Moreover, NQE improves robustness against noise. To validate the effectiveness of NQE, we conduct experiments on IBM quantum devices for image data classification, resulting in a remarkable accuracy enhancement from 0.52 to 0.96. In addition, numerical analyses highlight that NQE simultaneously improves the trainability and generalization performance of quantum neural networks, as well as of the quantum kernel method.
\end{abstract}
\maketitle
\section{Introduction}
\label{sec:intro}

Machine learning (ML) is ubiquitous in modern society, owing to its capability to identify patterns from data. For well-behaved data, simple learning algorithms such as linear regression and support vector machines are often sufficient to capture the underlying data distribution. In contrast, intricate and high-dimensional data typically require advanced learning algorithms, substantial computational power, and extensive training data. The overarching objective of machine learning is to construct models that can effectively learn the underlying distributions of complex real-world data. However, achieving this objective presents a substantial challenge.

Recent advances in quantum computing (QC) have led to the development of quantum machine learning (QML). QML aims to efficiently process complex data distributions by leveraging the computational benefits of quantum algorithms~\cite{Rebentrost_2018, PhysRevLett.113.130503_QSVM, lloyd2013quantum, QML-Biamonte,10.1038/s43588-022-00311-3}. 
One potential benefit of QC relevant to ML is its ability to efficiently sample from certain probability distributions that are exponentially difficult for classical counterparts~\cite{aaronson2011computational,Lund2017npjQI,HarrowMondanaro2017QCS}, as validated in several experiments~\cite{arute2019quantum,zhong2020quantum, madsen2022quantum}.
Quantum sampling algorithms typically impose lower requirements on physical implementations, making them an attractive pathway for demonstrating the quantum advantage using Noisy Intermediate-Scale Quantum (NISQ) devices~\cite{Preskill2018quantumcomputingin}.
One of the primary rationales underpinning the potential of QML is as follows: if a quantum computer can efficiently sample from a computationally hard probability distribution, it is plausible that quantum computers can efficiently learn from data drawn from such distributions. This implies a potential quantum advantage, especially for data distributions that are computationally infeasible for classical models but easily tractable for quantum models.

While quantum data is naturally suited for QML tasks~\cite{10.1038/s43588-022-00311-3}, most contemporary data science challenges involve classical data. Consequently, exploring the effectiveness of QML algorithms in learning from classical data constitutes a critical research focus. Notable examples of QML models tailored for classical data include Quantum Neural Network (QNN) and Quantum Kernel Method (QKM), both of which are specialized for supervised learning problems. QNN utilizes parameterized quantum circuit where the parameters are optimized through variational method~\cite{cong_quantum_2019, benedetti_parameterized_2019, abbas2021_a, cerezo2020variational}. In contrast, QKM utilizes quantum kernel function to effectively capture the correlations within the data~\cite{Havlicek2019,RigorousRobustQSpeedUp}.

In QML tasks involving classical data, an essential initial step is quantum embedding, which maps classical data into quantum states that a quantum computer can process. Quantum embedding is of paramount importance because it can significantly impact the performance of the learning model, including aspects such as expressibility~\cite{schuld2021effect}, generalization capability~\cite{caro2021encoding}, and trainability~\cite{thanasilp2022exponential}. Therefore, selecting an appropriate quantum embedding circuit is crucial for the successful learning of the data with quantum models. To achieve a quantum advantage in machine learning, prevailing research emphasizes on designing quantum embedding circuits that are computationally challenging to simulate classically~\cite{Havlicek2019, schuld2019quantum}. In this work, we redirect attention to the data separability of embedded quantum states, utilizing trace distance, a tool in quantum information theory used to measure the distinguishability between quantum states~\cite{Nielsen:2011:QCQ:1972505,wilde_2013}, as a figure of merit. Subsequent sections will show that the choice of a quantum embedding circuit inherently dictates a lower bound of empirical risk, independent of any succeeding trainable quantum circuits. Specifically, in the context of binary classification employing a linear loss function, the empirical risk is bounded from below by the trace distance between two ensembles of data-embedded quantum states representing different classes. Therefore, opting for a quantum embedding that maximizes trace distance---and thus, enhances distinguishability of states---facilitates improved training performance. Furthermore, a larger trace distance enhances resilience to noise, as the data-embedded quantum states reside farther from the decision boundary.

Conventional quantum embedding schemes are generally data-agnostic and do not guarantee high levels of data separability for a given dataset. To achieve a large trace distance, the use of trainable quantum embeddings is essential. Some efforts have explored trainable quantum embeddings by employing parameterized quantum circuits in both QNN~\cite{lloyd2020quantum} and QKM~\cite{PhysRevA.106.042431} frameworks. However, incorporating these quantum circuits increases the quantum circuit depth and the number of gates, making it less compatible with NISQ devices. Furthermore, the inclusion of trainable quantum gates during the quantum embedding phase increases the model's susceptibility to barren plateaus~\cite{10.1038/s41467-018-07090-4}, thereby adversely affecting the efficient training.

Given these considerations, we present Neural Quantum Embedding (NQE), an efficient method that leverages the power of classical neural networks to learn the optimal quantum embedding for a given problem. NQE can enhance the quantum data separability beyond the capabilities of quantum channels, thereby extending the fundamental limits of quantum supervised learning. Our approach avoids the critical issues present in existing methods, such as the increased  number of gates and quantum circuit depth and the exposure to the risk of barren plateaus. Numerical simulations and experiment with IBM quantum devices confirm the effectiveness of NQE in enhancing QML performance in several key metrics in machine learning. These improvements extend to training accuracy, generalization capability, trainability, and robustness against noise, surpassing the capabilities of existing quantum embedding methods.

\section{Results}

\subsection{Lower Bound of Empirical Risk in Quantum Binary Classification}
\label{sec:qml}

In supervised learning, the primary objective is to identify a prediction function $f$ that minimizes the true (expected) risk $R(f) = \mathbb{E}[l(f(X),Y)]$ with respect to some loss function $l$, where $X$ and $Y$ are drawn from an unknown distribution $D$. Given a collection of $N$ sample data $\lbrace (x_{i}, y_{i})\rbrace$, the goal of learning algorithms is to find the optimal function $f^{*}$ that minimizes the empirical risk $R_N(f)=(1/N)\sum_{i=1}^{N} l(f(x_{i}), y_{i})$ among a fixed function class $F$, i.e., $f^{*} = \arg\min_{f \in F} R_N(f)$. Quantum supervised learning algorithms aim to efficiently find prediction functions with improved performance by exploiting the computational power of the quantum device.

A QNN is a widely used method for quantum supervised learning. In QNN, a classical input data $x$ is first embedded into a quantum state by applying a quantum embedding circuit to an initial ground state, resulting in $\ket{x} = \Phi(x)\ket{0}^{\otimes n}$. Next, a parameterized unitary operator, denoted as $U(\theta)$, is applied to transform the embedded quantum states, and the state is measured with an observable $O$. The measurement outcome serves as a prediction function for supervised learning algorithms, expressed as $f(x; \theta) = \bra{x} U^{\dagger}(\theta) O U(\theta)\ket{x}$ . Subsequently, using gradient descent or one of its variants, we search for the optimal parameter $\theta^{*}$ that minimizes the empirical risk. For a binary classification task with input data $x\in\mathbb{R}^{m}$ and its associated label $y \in \{-1, 1\}$, we can predict the label of the new data $x_\text{new}$ using the rule $y_\text{new} = \text{sign}[f(x_\text{new}; \theta^{*})]$.

Alternatively, we can consider this procedure as a quantum state discrimination problem involving two parameterized positive operator-valued measures (POVMs), denoted as $E_{\pm}(\theta) = (I \pm U^{\dagger}(\theta) O U(\theta))/2$. With these POVMs, the probabilities of obtaining measurement outcomes $\pm 1$ given an input data $x$ are computed as $P(E_{\pm}(\theta) \vert x)=\bra{x}E_{\pm}(\theta)\ket{x}$. Subsequently, the decision rule for the new data is determined as $y_\text{new} = \text{sign}[P(E_{+}(\theta^{}) \vert x_\text{new}) - P(E_{-}(\theta^{}) \vert x_\text{new})]$.
In such a scenario, a natural loss function is the probability of misclassification, which can be expressed as $l(f(x; \theta), y) = P(E_{\neg y}(\theta) \vert x)$. Considering a dataset of $N$ samples $S = \{x_i^{-}, -1\}_{i=1}^{N^{-}} \cup \{x_i^{+}, 1\}_{i=1}^{N^{+}}$, the empirical risk becomes
\begin{align}
\label{eq:loss}
    L_s &= \frac{1}{N} \bigg{[}\sum_{i=1}^{N^{-}}P(E_{+}(\theta) \vert x^{-}_{i}) + \sum_{i=1}^{N^{+}}P(E_{-}(\theta) \vert x^{+}_{i}) \bigg{]} \nonumber \\
    &\geq \frac{1}{2} - D_{\text{tr}}(p^{-}\rho^-, p^{+}\rho^+),
\end{align}
where $ \rho^{\pm} = \sum \ket{x_i^{\pm}}\bra{x_i^{\pm}}/N^{\pm}$, $p^{\pm} = N^{\pm}/N$, and $D_{\text{tr}}(\cdot\;,\cdot)$ denotes the trace distance~\cite{Bae_2015}. It is important to note the contractive property of the trace distance given by
\begin{equation}
\label{eq:contractive}
    D_{\text{tr}}(\Lambda(\rho_0),\Lambda(\rho_1))\le D_{\text{tr}}(\rho_0,\rho_1),
\end{equation}
for any positive and trace-preserving (PTP) map $\Lambda$~\cite{e23050625}.
Based on the above, we now emphasize two crucial points.
\begin{enumerate}
    \item The empirical risk is lower bounded by the trace distance between two data ensembles $p^{-}\rho^-$ and $p^{+}\rho^+$. This bound is completely determined by the initial quantum embedding circuit, regardless of the structure of the parameterized unitary gates $U(\theta)$ applied afterwards.
    \item The minimum loss is achieved when $\lbrace E_{-}(\theta),E_{+}(\theta)\rbrace$ is a Helstrom measurement. Therefore, the training of a quantum neural network can be viewed as a process of finding the Helstrom measurement that optimally discriminates between the two data ensembles.
\end{enumerate}
Designing a quantum embedding that maximizes the trace distance is of paramount importance since it minimizes the lower bound of the empirical risk. This becomes especially important in NISQ applications, as non-unitary quantum operations, such as noise, strictly reduce the trace distance between two quantum states~\cite{Nielsen:2011:QCQ:1972505,wilde_2013}. Therefore, there is a clear need for a trainable, data-dependent embedding that can maximize the trace distance.

Several works have proposed combining a set of parameterized quantum gates and a conventional quantum embedding circuit as a means to create a trainable unitary embedding~\cite{lloyd2020quantum,glick2021covariant,PhysRevA.106.042431}. However, the use of parameterized quantum gates comes with several drawbacks. Firstly, it results in an increase in the number of gates and the depth of the quantum circuit. This not only increases computational costs but also makes the quantum embedding more susceptible to noise. Furthermore, the method is prone to encountering barren plateaus, which pose a fundamental obstacle to scalability~\cite{10.1038/s41467-018-07090-4,thanasilp2022exponential}. Secondly, the trainable unitary embedding is highly restricted in enhancing the maximum trace distance of embedded quantum states (see Section~\ref{sec:NQEvsTUE}). It is crucial to note that none of the existing quantum embeddings can guarantee the effective separation of two data ensembles in the Hilbert space with a large distance.

\subsection{Neural Quantum Embedding}
\label{sec:NQE}
\begin{figure*}[ht]
    \centering
    \includegraphics[width=0.7\textwidth]{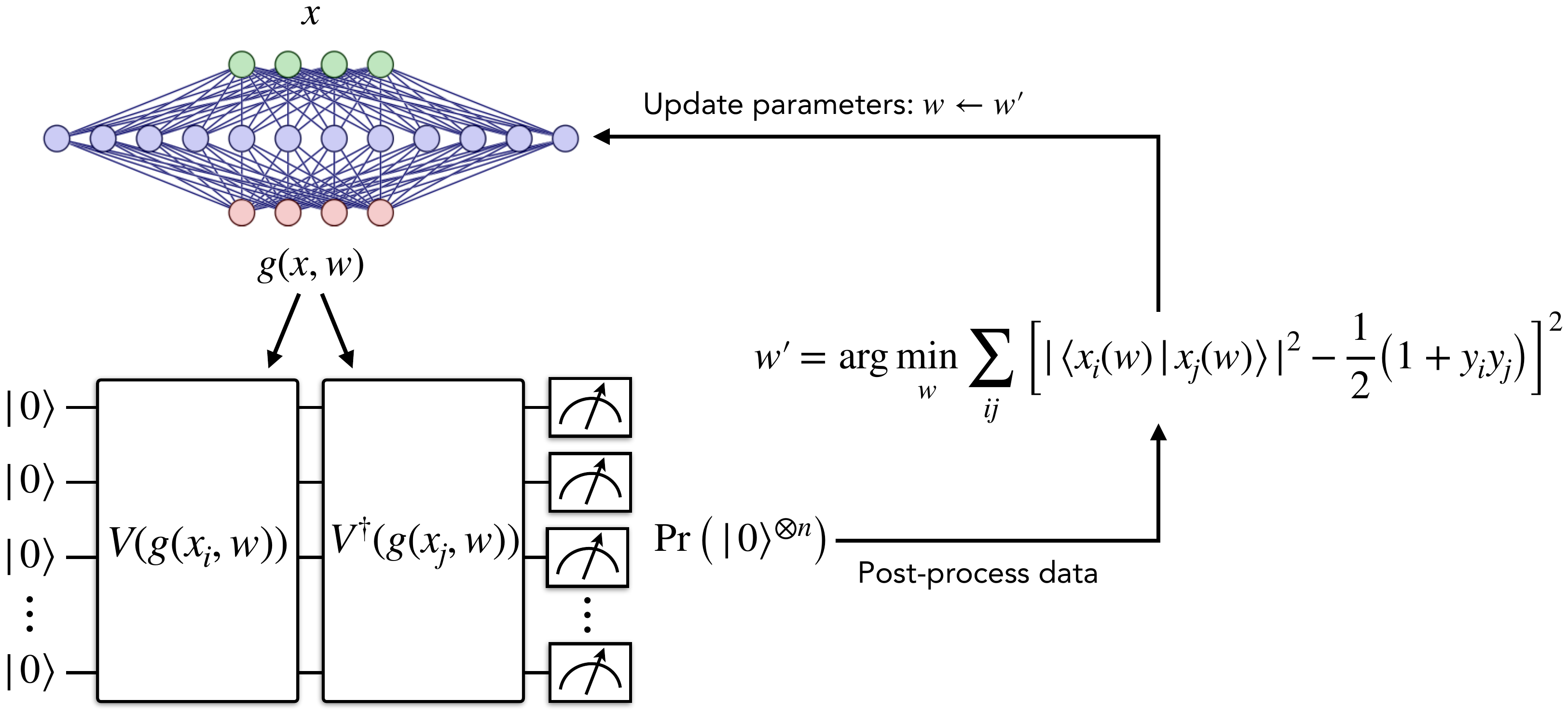}
    \caption{Overview of the NQE training. The unitary transformation that maps $x_i$ to the quantum feature space is determined by the output of a classical neural network denoted by $g(x_i,w)$, where $w$ represents trainable parameters. The resulting quantum state is $|x_i(w)\rangle = V(g(x_i,w))|0\rangle^{\otimes n}$. The goal of the training is to produce mapping functions that can separate the two classes of data into two orthogonal subspaces. Efficient calculation of the fidelity between the two quantum states produced by the feature map is performed using a quantum computer.}
    \label{fig:feature_map_training_fidelity}
\end{figure*}

Neural Quantum Embedding utilizes a classical neural network to maximize the trace distance between two ensembles $D_{\text{tr}}(p^{-}\rho^{-}, p^{+}\rho^{+})$. It can be expressed as $\Phi_{\mathrm{NQE}}: x \rightarrow \ket{x} = V(g(x,w))\ket{0}^{\otimes n}$, where $V$ is a general quantum embedding circuit and $g: \mathbb{R}^m\times\mathbb{R}^{r} \rightarrow \mathbb{R}^{m'}$ is a classical neural network that transforms the input data $x$ using $r$ trainable parameters.

By choosing $m' < m$, we can bypass additional classical feature reduction methods, such as principal component analysis (PCA) or autoencoders, typically employed prior to quantum embedding due to the current limitations on the number of reliably controllable qubits in quantum devices.
Ideally, the loss function should directly contain the trace distance. However, calculating it is computationally expensive even with the quantum computer. Therefore, we used an implicit loss function derived from a fidelity measure, which is expressed as
\begin{equation}
\label{eq:loss}
    l_{\text{fid}}((x_i, y_i), (x_j,y_j)) = \left[ \left\vert \braket{x_i}{x_j} \right\vert^2 - \frac{1}{2}\left(1+y_{i} y_{j}\right)\right]^2.
\end{equation}
This fidelity loss can be efficiently computed using the swap test~\cite{PhysRevLett.87.167902} or directly measuring the state overlap (see Figure~\ref{fig:feature_map_training_fidelity}). The relationship between the state fidelity and the trace distance, as well as how minimizing $l_{\text{fid}}$ corresponds to enhancing the trace distance are detailed in Appendix~\ref{Appendix:C}.

While NQE is not restricted by the choice of the quantum embedding circuit, we specifically focus on improving the ZZ feature embedding~\cite{Havlicek2019}. 
The unitary operator corresponding to this embedding is expressed as
\begin{equation}
\label{eq:qembed}
     V(\phi(x)) = \bigg{[} \exp\bigg{(}i \sum_i \phi_i(x)Z_i  + i \sum_{i,j}\phi_{i,j}(x) Z_i Z_j \bigg{)} H^{\otimes n}\bigg{]}^{L},
\end{equation}
where $L>1$.
The use of this embedding is prevalent due to the conjectured intractability of computing its kernel classically when $L>1$~\cite{Havlicek2019}. It has been extensively explored in the field of quantum machine learning, including theoretical investigations~\cite{abbas2021_a,huang_power_2021,VQASVM} as well as practical applications in areas like drug discovery~\cite{doi:10.1021/acs.jcim.1c00166,Mensa_2023}, high energy physics~\cite{PhysRevResearch.3.033221,Li_2023}, and finance~\cite{9643469,9915517}. The most commonly used functions for $\phi$ are $\phi_i(x) = x_i$ and $\phi_{i,j}(x) = (\pi - x_i)(\pi - x_j)/2$~\cite{Havlicek2019, abbas2021_a}, but these choices are made without justifications. Although  Ref.~\cite{suzuki_analysis_2020} numerically illustrates that the choice of $\phi$ can significantly impact the performance of QML algorithms, it does not provide guidelines for selecting an appropriate $\phi$ for the problem at hand. NQE effectively solves this limitation by replacing mapping functions with a trainable classical neural network.

While optimizing the ZZ feature embedding through NQE requires the use of a quantum computer due to the computational hardness of the loss function, there are certain embeddings that can be optimized solely on a classical computer. An instance of this is the amplitude encoding~\cite{PhysRevA.102.032420,PhysRevA.101.032308,9259210,araujo_divide-and-conquer_2021, araujo_configurable_2023}, where the corresponding loss function for NQE can be computed by taking the dot product of two vectors. 
\subsection{Experimental Results}
\label{sec:exp}

\subsubsection{NQE versus Fixed Unitary Embedding}
\label{sec:NQEvsFUE}
This section presents experimental results that demonstrate the effectiveness of NQE in enhancing QML algorithms. To this end, we employed a four-qubit Quantum Convolutional Neural Network (QCNN)~\cite{cong_quantum_2019,pesah2020absence,hur2022quantum,kim2023classical,oh2023quantum} for the task of classifying images of 0 and 1 within the MNIST dataset~\cite{lecun2010mnist}, a well-established repository of handwritten digits. We conducted experiments using both noiseless simulations and quantum devices accessible through the IBM cloud service. The experiment unfolds in three main phases: the application of NQE, the training of the QCNN with and without NQE models, and the assessment of classification accuracies for the trained QCNN with and without NQE models.

\begin{figure*}[ht]
    \centering
        \includegraphics[width=0.96\textwidth]{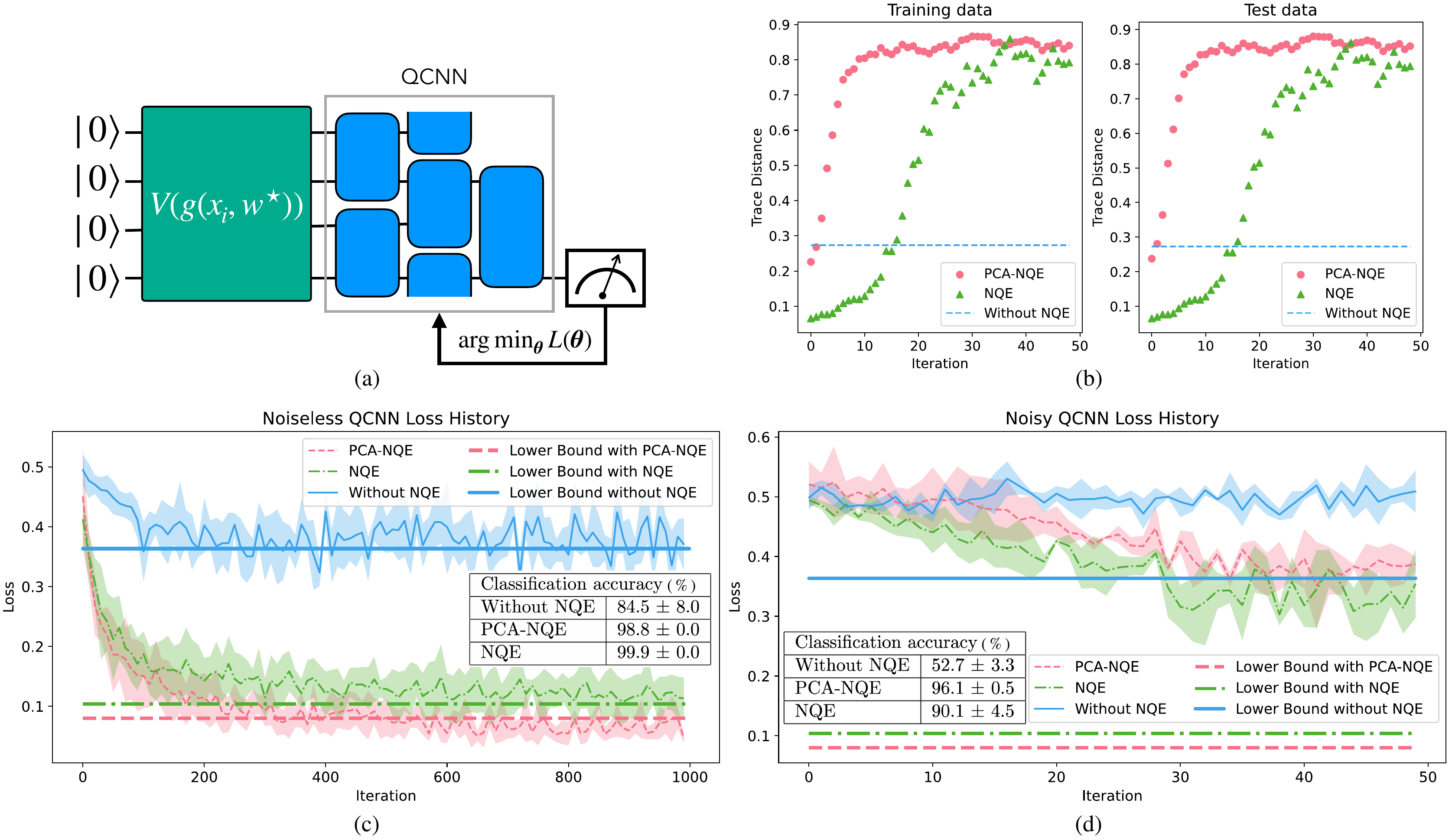}
        \caption{(a) Schematic representation of the quantum circuit used in the experiments. The green rectangle indicates the Neural Quantum Embedding (NQE), which transforms classical data $x_i$ into a quantum state $\vert x_i \rangle$. The blue rectangles represent two-qubit parameterized quantum gates of the Quantum Convolutional Neural Network (QCNN), designed for binary classification tasks. (b) Plot depicting the evolution of the trace distance between two ensembles of quantum states embedded by the NQE models during training on the ibmq\_toronto device, compared to the trace distance from conventional quantum embedding without NQE. (c) Noiseless QCNN simulation results. (d) The results from QCNN experiments conducted on IBM quantum devices. In (c) and (d), the blue solid, red dashed, and green dash-dotted lines represent the mean training loss histories for conventional ZZ feature embedding, PCA-NQE, and NQE, respectively. The shaded regions in the figure represent one standard deviation from the mean. These values are acquired from five repetitions of each QCNN training with random initialization of parameters. The thicker versions of these lines indicate the theoretical lower bounds for each method.
        }
        \label{fig:qcnn_demonstration}
\end{figure*}

We compared NQE against the conventional ZZ feature embedding with the aforementioned function $\phi_i(x)=x_i$ and $\phi_{i,j}(x)=(\pi - x_i)(\pi - x_j)/2$. Due to the limited number of qubits that can be manipulated reliably in current quantum devices, it is often necessary to reduce the number of features in the original data before embedding it as a quantum state. To address this issue, we tested two different NQE structures for incorporating dimensionality reduction. In the first approach, which we refer to as PCA-NQE, we applied PCA to reduce the number of features before passing them to the neural network. In the second approach, which we simply refer to as NQE, dimensionality reduction was directly handled within the neural network by adjusting the number of input and output nodes accordingly. For both PCA-NQE and NQE, classical neural networks produce eight dimensional vectors, which are then used as rotational angles for the four-qubit ZZ feature embedding. These two methods differ in their input requirements: PCA-NQE takes four-dimensional vectors as input, requiring classical feature reduction, while NQE accepts the original 28-by-28 image data, bypassing the need for additional classical preprocessing. Further details on the structure of the ZZ feature embedding circuit and the classical neural networks used for NQE methods are provided in Appendix~\ref{appendix:1}.

The goal of NQE training is to maximize the trace distance between the two sets of data ensembles, thereby minimizing the lower bound of the empirical risk. Figure~\ref{fig:qcnn_demonstration}(b) displays the trace distance as a function of NQE training iterations, acquired from ibmq\_toronto. It is evident that both PCA-NQE and NQE effectively separate the embedded data ensembles and enhance the distinguishability of quantum state representing different classes, even in the presence of noise in the real quantum hardware. After optimization, PCA-NQE and NQE yield trace distances of 0.840 and 0.792, respectively, which are significantly improved compared to the conventional ZZ feature embedding with a distance of 0.273. Consequently, the lower bound of the empirical risk is significantly reduced to 0.08 and 0.104, respectively, from the original value of 0.364.

The training of the QCNN circuit provides additional evidence supporting the effectiveness of NQE in QML. Figure~\ref{fig:qcnn_demonstration}(c) presents the results of noiseless QCNN simulations. In this figure, 
the blue solid, red dashed, green dash-dotted lines represent the mean training loss histories for conventional ZZ feature embedding, PCA-NQE, and NQE, respectively. The thicker versions of these lines indicate the theoretical lower bounds for each method. The mean values are calculated from five repetitions of each QCNN training with random initialization of parameters. The shaded regions in the figure illustrate one standard deviation from the mean. The empirical risks for all models converge toward their respective theoretical minima, affirming that the trained QCNN adequately approximates optimal Helstrom measurements. Classification accuracies achieved on the test dataset are shown in the table within the figure. NQE methods led to significant enhancements, as demonstrated by reduced empirical risks and improved classification accuracies. Improvements in classification accuracies are expected as NQE embeds the training data into a specific subspace within the given Hilbert space such that the state distinguishability is maximized. The localization of embedded data facilitates the use of ML models with reduced complexity, thereby implying an enhancement in generalization capability. Numerical results supporting this intuition are presented in Sections~\ref{sec:results_ed}, \ref{sec:QKM}, and Appendix~\ref{appendix:expressibility}. Note that in some instances, the training loss falls below the theoretical lower bound of empirical risk. This occurs because the training loss is computed from a randomly sampled mini-batch of data in each iteration.

Figure~\ref{fig:qcnn_demonstration}(d) presents the mean QCNN training loss histories obtained using IBM quantum devices. The mean values are calculated from three independent trials on ibmq\_jakarta, ibmq\_perth, and ibmq\_toronto devices with random initialization of parameters. The shaded regions in the figure illustrate one standard deviation from the mean. The presence of noise and imperfections in the quantum devices prevents the empirical risks from reaching their theoretical lower bounds. Nonetheless, the training performance is significantly improved by NQE. Notably, for both NQE methods, the empirical risk rapidly approaches or even falls below the theoretical limit of the conventional method. This result underscores that, even on the current noisy quantum devices, our method can outperform the theoretical optimum of the ZZ feature map without any additional quantum error mitigation. Moreover, a substantial improvement in classification accuracy was recorded, reaching 96\% (PCA-NQE) and 90\% (NQE), whereas the conventional method achieved only 52\%. These findings demonstrate that NQE enhances the noise resilience of QML models, improving the utility of NISQ devices. Comprehensive details of these experiments are provided in Appendix~\ref{appendix:2}.

\subsubsection{NQE versus Trainable Unitary Embedding}
\label{sec:NQEvsTUE}
\begin{figure*}[ht]
    \centering
        \includegraphics[width=0.96\textwidth]{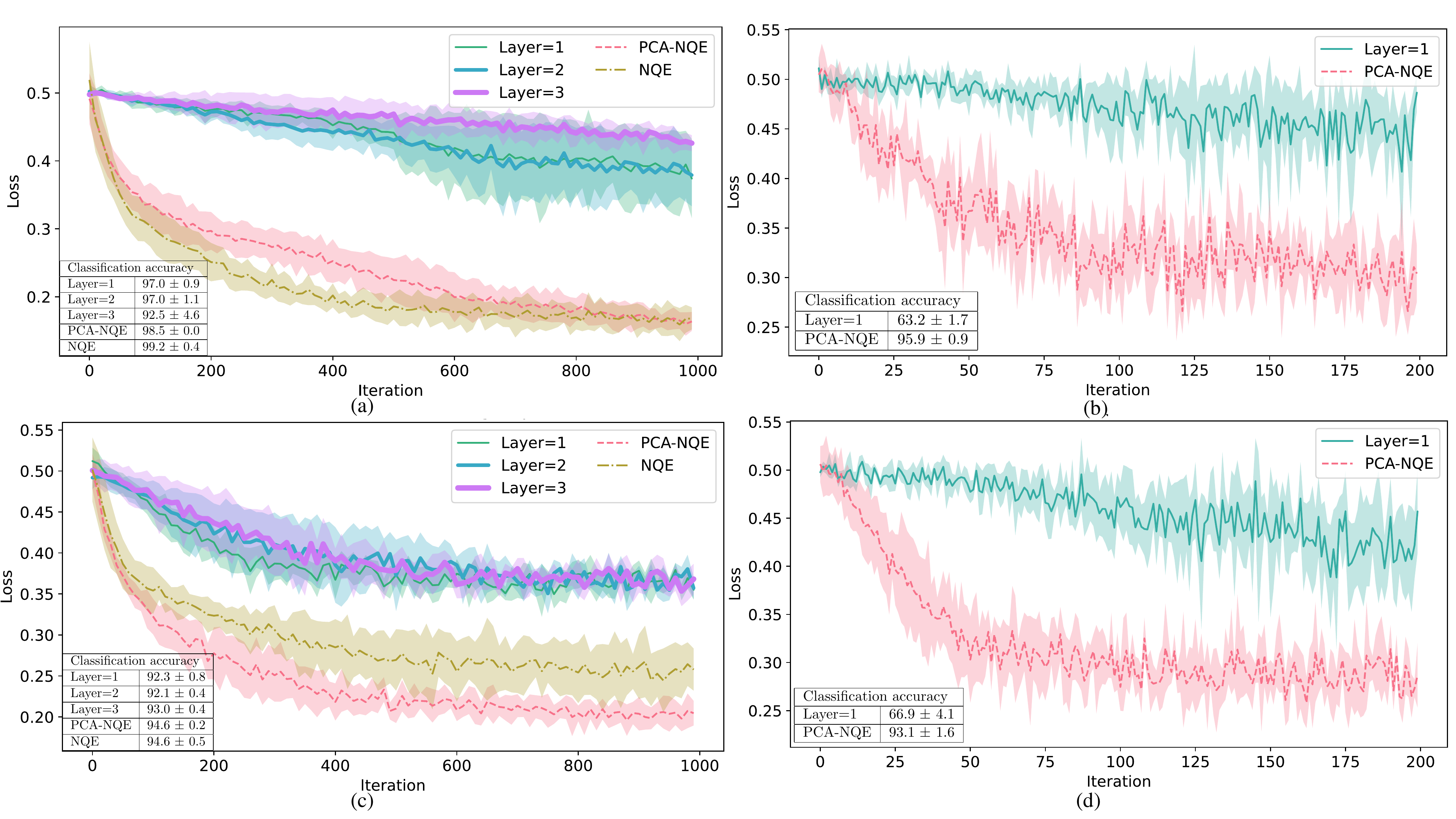}
        \caption{Comparative analysis between Neural Quantum Embeddings and Trainable Unitary Embeddings with one, two and three trainable layers. The numerical simulations were conducted under noiseless (left) and noisy (right) environments, utilizing MNIST (top) and Fashion-MNIST (bottom) datasets. For noiseless simulations, we used 1000 iterations, a learning rate of 0.01 learning rate, and batches of 128 data points per iteration. For noisy simulations, we used 200 iterations, a learning rate of 0.05, and batches of 15 data point per iteration. The noisy model simulations utilized the IBM Qiskit FakeGuadalupe environment. The classification accuracies were evaluated using a sample size of 2115 and 2000 data points for the MNIST and Fashion-MNIST datasets, respectively. The mean and one standard deviation from five independent iterations are shown for the loss history.}
        \label{fig:NQEvsTUE}
\end{figure*}

We also conduct numerical comparisons between NQE and trainable unitary embedding to further investigate the advantage of NQE.
Trainable unitary embedding utilizes trainable unitary $U_\mathrm{tra}(\theta)$ to find the quantum embedding that separates the data well. More specifically, one can implement a trainable unitary embedding $\Phi(x;\theta)$ as
\begin{equation}
    \Phi(x;\theta): \vert 0 \rangle ^{\otimes n} \rightarrow \vert x;\theta \rangle = U_\mathrm{emb}(x)U_\mathrm{tra}(\theta)\vert 0 \rangle^{\otimes n}.
\end{equation}
In this case, the data embedded ensembles are expressed as,
\begin{align}
\rho^\pm(\theta) &=\frac{1}{N^\pm}\sum_{i} U_\mathrm{emb}(x^\pm_i) U_\mathrm{tra}(\theta)\vert 0 \rangle \langle 0 \vert ^{\otimes n} U_\mathrm{tra}^\dagger(\theta)U_\mathrm{emb}^\dagger (x_i^\pm) \nonumber
\\
&= \mathcal{E}^{\pm}(\vert \psi (\theta) \rangle \langle \psi(\theta) \vert),
\end{align}
where $\vert \psi (\theta) \rangle = U_\mathrm{tra}(\theta)\vert 0 \rangle^{\otimes n}$ and $\mathcal{E}^{\pm}(\cdot)$ are quantum channels that maps $\rho \rightarrow \sum_{i} K^{\pm}_i \rho K_i^{\pm\dagger}\nonumber $ with $K^\pm_i = U_\mathrm{emb}(x_i^\pm) / \sqrt{N^\pm}$. Now the maximum trace distance between two data ensembles is upper bounded by the diamond distance, 
\begin{align}
    &\max_\theta D_\text{tr}(p^+\rho^+(\theta), p^-\rho^-(\theta)) =
    \nonumber \\
    &\max_{\theta} \Vert p^+\mathcal{E}^+ (\vert \psi (\theta) \rangle \langle \psi (\theta) \vert) - p^-\mathcal{E}^- (\vert \psi (\theta) \rangle \langle \psi (\theta) \vert) \Vert_1  \nonumber\\
    & \leq D_\diamond(p^+\mathcal{E}^+, p^-\mathcal{E}^-).
\end{align}
This presents a significant limitation since $\mathcal{E}^{\pm}$ are entirely predetermined by the choice of quantum embedding circuit $U_{\text{emb}}(\cdot)$, without any guarantee that the diamond distance will be large. The trainable unitary $U_\mathrm{tra}(\theta)$ does not contribute to improving the upper bound of the maximum trace distance.

Alternatively, one may consider the data re-uploading technique in which the trainable unitary and quantum embedding circuit are repeatedly applied multiple times:
\begin{equation}
    \Phi(x;\theta): \vert 0 \rangle ^{\otimes n} \rightarrow \vert x;\theta \rangle = \prod_{l=1}^{L}U_\text{emb}(x)U_\text{tra}(\theta_l)\vert 0 \rangle^{\otimes n}.
\end{equation}
However, Ref~\cite{jerbi_quantum_2023} demonstrates that data-reuploading quantum embedding can be exactly transformed into a form where all the trainable unitaries follow the quantum embedding circuits by introducing ancilla qubits. Upon such transformation, the embedding can be expressed as
\begin{equation}
    U'_\mathrm{tra}(\boldsymbol{\theta}) U'_\mathrm{emb}(x)\vert 0 \rangle^{\otimes n+n'},
\end{equation}
where $\boldsymbol{\theta}=\lbrack \theta_1,\ldots,\theta_L\rbrack$ and $n'\in O(L\log(L))$. Importantly, the embedding circuit is independent of the parameters of the trainable unitary gates. Consequently, the maximum trace distance is once again determined by the choice of quantum embedding circuit, which, in turn constrains the data distinguishability. Furthermore, employing multiple layers of trainable unitary and quantum embedding circuit increases the total circuit depth significantly, making it not only unsuitable for NISQ applications but also susceptible to the barren plateaus problem.

Figure~\ref{fig:NQEvsTUE} displays a comparison of the NQE and trainable unitary embedding under noiseless and noisy environments employing eight qubits. The PCA-NQE and NQE are implemented with adjustments made in both PCA and the classical neural network to accommodate the use of eight qubits (see Appendix~\ref{appendix:1}). The QCNN circuits are trained utilizing either trainable unitary embedding or NQE techniques. For the trainable unitary embedding we used following parameterized quantum circuit,
\begin{equation}
\prod_{l=1}^{L}\left[V(\phi(x))\exp\left(i\sum_i \theta^l_i Y_i + i \sum_{i,j} \theta^l_{i,j}Y_iY_j\right)\right] \vert 0 \rangle^{\otimes n}.
\end{equation}
Here, $V(\phi(x))$ is the ZZ feature map explained in Eq.~(\ref{eq:qembed}) and $\theta^l_i\;(\theta^l_{i,j})$ are the trainable parameters of the embedding.

Figure~\ref{fig:NQEvsTUE} (a) and (c) depict the training loss history and classification accuracies in a noiseless environment for the MNIST ($\{0,1\}$) and Fashion-MNIST ($\{\text{T-shirt/Top},\text{Trouser}\}$) datasets~\cite{xiao2017/online}, respectively. In this experiment, we explored the impact of one, two, and three trainable unitary layers in comparison to PCA-NQE and NQE. The results indicate that the NQE methods achieves a notably lower training loss than the trainable unitary embedding, suggesting superior efficacy of NQE in improving data separability (by increasing the trace distance). Additionally, the NQE methods resulted in higher classification accuracies.

Figure~\ref{fig:NQEvsTUE}, panels (b) and (d), depict the training loss history and classification accuracies in a noisy environment for the MNIST and Fashion-MNIST datasets, respectively. A single-layer trainable unitary embedding was evaluated against PCA-NQE. The choice of a single layer was based on its minimal circuit depth, rendering it less susceptible to noise interference. The results indicate NQE are more effective in enhancing data separability under presence of noise. Additionally, PCA-NQE yields considerably higher classification accuracies. These advantages stem from the larger initial trace distance and the reduced circuit depths associated with PCA-NQE. We employed a noisy simulation using the IBM FakeGuadalupe device. This simulator mimics the essential characteristics of the ibmq\_guadalupe device, including its basis gates, qubit connectivity, qubit relaxation (T$_1$) and rephrasing (T$_2$) times, and readout error rates.
\subsection{Effective Dimension}
\label{sec:results_ed}

As demonstrated thus far, NQE has proven to be an efficient method for reducing the lower bound of empirical risk. While this advancement enhances the ability to learn from data, another essential measure of successful machine learning is the generalization performance, that is, the ability to make accurate predictions on unseen data based on what has been learned. The simulation and experimental results presented in the previous section indicate an improvement in prediction accuracy on test data with the implementation of NQE. In this section, we provide additional evidence of improved generalization performance by analyzing the effective dimension (ED)~\cite{berezniuk2020scale, abbas2021_a, abbas2021_b} of a QML model constructed with and without NQE. Intuitively, ED quantifies the number of parameters that active in the sense that they influence the outcome of its statistical model. 

\begin{figure}[t]
    \centering
    \includegraphics[width=0.95\columnwidth]{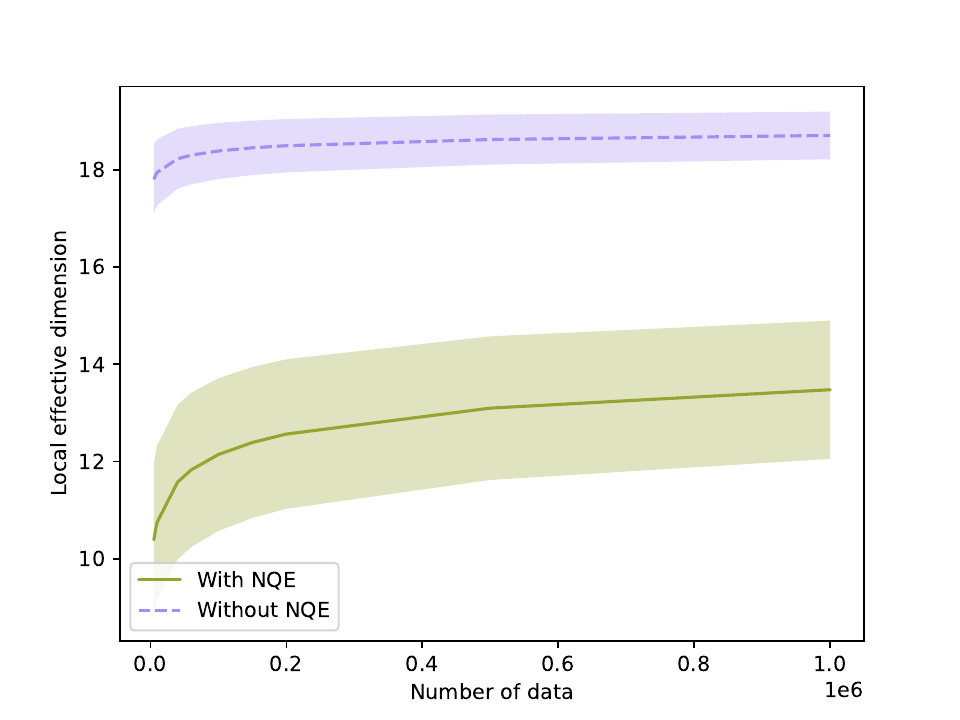}
    \caption{The local effective dimension for the circuit with (solid green) and without (dashed blue) NQE. These results are based on ten sets of experiments, each on a distinct artificial dataset with 20 repetitions with random initialization of parameters. The reported values represent the average across all 200 experiments.} \label{fig:effective_dimension}
\end{figure}

To investigate this further, we focus on the local effective dimension (LED) introduced in Ref.~\cite{abbas2021_b}, as it takes into account the data and learning algorithm dependencies and is computationally more convenient. Importantly, the LED exhibits a positive correlation with the generalization error, allowing for straightforward interpretation of the results: a smaller LED corresponds to a smaller generalization error, and vice versa. Our numerical investigation employs a four-qubit QNN (see Appendix~\ref{appendix:3}), and the results are presented in Fig.~\ref{fig:effective_dimension}.  In the figure, the green solid and purple dashed lines represent the mean local effective dimension of the tested QML model with and without NQE, respectively. The mean values are computed from 200 trials, where each trial consists of ten artificial datasets, and for each dataset, the experiment is repeated 20 times with random initialization of parameters. The shaded areas in the figure represent one standard deviation. The simulation results unequivocally demonstrate that NQE consistently reduces the LED in all 200 instances tested and across a wide range of training data sizes, signifying an improvement in generalization performance.

The effective dimension can also be interpreted as a measure of the volume of the solution space that a specific model class can encompass. A smaller effective dimension in a QML model implies a reduced volume of the solution space. This observation is particularly relevant as it suggests that models with smaller effective dimensions are less prone to encountering barren plateaus~\cite{holmes_2022}. Consequently, the simulation results further indicate that NQE not only enhances generalization performance but also improves the trainability of the model. Further evidence substantiating this improvement for both QNN and QKM will be presented in a later section. 

\subsection{Generalization in Quantum Kernel Method}
\label{sec:QKM}
Up to this point, the investigation of NQE has primarily focused on its application within the context of quantum neural networks. In this section, we extend the analysis to demonstrate that NQE also enhances the performance of the quantum kernel method. Given a quantum embedding, the kernel function can be defined as
\begin{equation} \label{eq:quantum_kernel_matrix}
    k^Q(x_i, x_j) = \left\vert \braket{x_i}{x_j} \right\vert^2,
\end{equation}
which can be computed efficiently on a quantum computer. The quantum kernel method refers to an approach that uses the kernel matrix $K^Q$, of which each entry is the kernel of the corresponding data points, in a method like a classical support vector machine~\cite{Havlicek2019,huang_power_2021}. The potential quantum advantage of such approach is based on the hardness to compute certain quantum kernel functions classically~\cite{Havlicek2019,PhysRevLett.117.080501,RigorousRobustQSpeedUp}. 

The quantum kernel method attempts to determine the function $f(x;W) =  \text{Tr}(W \vert x \rangle \langle x \vert)$ to predict the true underlying function $h(x)$ for unseen data $x$. The optimal parameters $W^{*}$ are obtained by minimizing the cost function
\begin{equation}
W^{*} = \argmin_{W \in \mathbb{C}^{2^n \times 2^n}} \frac{1}{N} \sum_{i=1}^N\left( f(x_i;W) - h(x_i)
\right)^2 - \lambda \vert\vert W \vert\vert_{\text{F}}^2 ,
\end{equation}
where $\vert\vert \cdot \vert\vert_{\text{F}}$ is the Frobenius norm. The second term is the regularization term with a hyperparameter $\lambda$. The purpose of including the regularization term is to reduce the generalization error at the expense of the training error. Specifically, considering the true error $R(W)= \mathbb{E}_x \vert f(x;W) - h(x) \vert$, and the training error $R_N(W) = \sum_{i=1}^N \vert f(x_i;W) - h(x_i) \vert/N$, the generalization error is upper-bounded as
\begin{equation}
    \left \vert R(W) - R_N(W) \right \vert 
    \leq \mathcal{O}\left(\frac{\vert\vert W \vert\vert_{\text{F}}}{\sqrt{N}} + \sqrt{\frac{\log(1 / \delta)}{N}} \right),
\end{equation}
with probability at least $1-\delta$ (see Ref~\cite{huang_power_2021} Supplementary Information Section 4.C).
The optimal $W^{*}$ can be expressed as $W^{*} = \sum_{i=1}^N\sum_{j=1}^N h(x_i)(K^Q + \lambda I)^{-1}_{i,j} \vert x_j \rangle \langle x_j \vert$. Here, the employment of NQE affects $W^{*}$ as both $K^Q$ and $\vert x_j \rangle$ vary with the quantum embedding. 

We performed empirical evaluations to assess the effectiveness of NQE in reducing $G = \vert\vert W^* \vert\vert_{\text{F}}/\sqrt{N}$ and thereby improving the upper bound of the generalization error. The analysis proceeded in three steps: loading the dataset, computing the quantum kernel matrix with and without NQE, and calculating the generalization error bound with and without NQE. The experiments tested both PCA-NQE and NQE, with neural network parameters optimized on the ibmq\_toronto processor, as detailed in Appendix~\ref{appendix:1}. During the dataset loading phase, binary datasets containing $N =1,000$ samples from classes \{0,1\} were constructed from the MNIST dataset. As outlined in Appendix~\ref{appendix:1}, both PCA-NQE and the conventional quantum embedding without NQE were preceded by PCA to reduce the number of features to four. In contrast, the NQE utilized the original 28x28 image datasets. Subsequently, three quantum kernel matrices were constructed: one without NQE, one with PCA-NQE, and one with NQE. The $(i,j)$th entry of the quantum kernel corresponds to $k^Q(x_i,x_j)$, the fidelity overlap between pairs of data-embedded quantum states. This fidelity overlap was computed using Pennylane~\cite{bergholm2020pennylane} numerical simulation. Finally, for each of the quantum kernel matrices, the corresponding upper bound of the generalization error $G=\sqrt{\vert\vert W^* \vert\vert^2_{\text{F}}/N}$ was calculated. Here, $\vert\vert W^* \vert\vert^2_{\text{F}}$ can be expressed as $\vert\vert W^* \vert\vert^2_{\text{F}}=\sum_{i} \sum_{j} \left[(K^Q + \lambda I)^{-1}K^Q(K^Q + \lambda I)^{-1}\right]_{i,j}y_iy_j$.  

\begin{figure}[t]
    \centering
    \includegraphics[width=1.0\columnwidth]{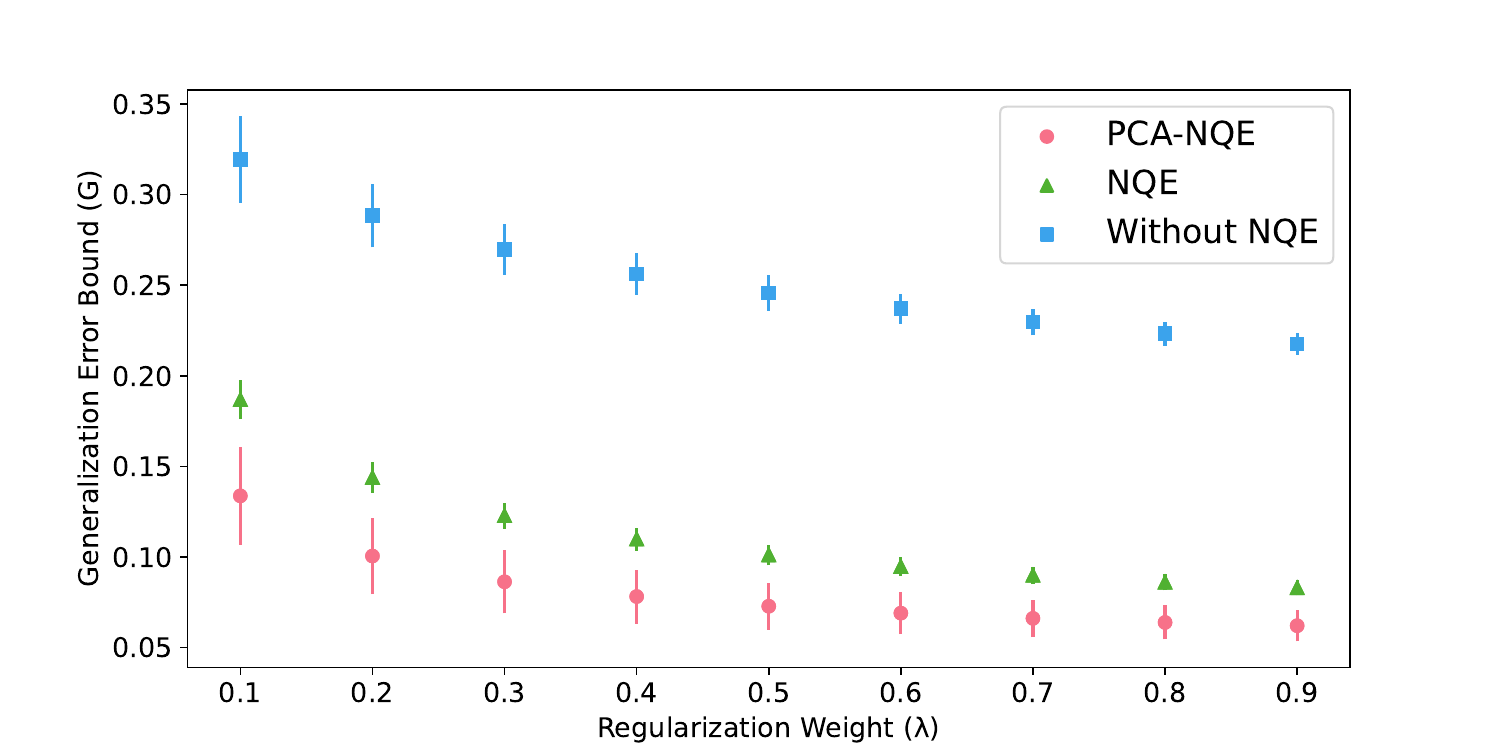}
    \caption{
    A comparative analysis of the generalization error bound $G$ with varying regularization weights $\lambda$. This plot illustrates the performance enhancement -- lower generalization error bound -- when employing NQE (green triangles) and PCA-NQE (red circles) over conventional methods without NQE (blue squares) in quantum kernel methods. PCA-NQE and NQE were optimized on the ibmq\_toronto quantum hardware. The error bound $G$ was determined based on five independent numerical simulations, presenting both the mean and one standard deviation of $G$.
    } \label{fig:QKM Generalization}
\end{figure}

The experimental procedure was repeated five times, each time with different sets of 1,000 samples of data. Figure~\ref{fig:QKM Generalization} presents the mean and one standard deviation of the generalization error bound $G$. These values were obtained under the ZZ feature embedding, both with and without NQE, and were examined across various regularization parameters $\lambda$. The results clearly illustrate that NQE significantly decreases the upper bound of the generalization error in quantum kernel methods.


\subsection{Expressibility and Trainability}
\label{sec:ExpTrain}
In both QNN and QKM, there exists a trade-off between expressibility and trainability. In the QNN framework, highly expressive quantum circuits often leads to barren plateaus, characterized by exponentially vanishing gradients, which severely hinders the trainability of the model~\cite{10.1038/s41467-018-07090-4, holmes_2022}. In the QKM framework, highly expressive embedding induces a quantum kernel matrix whose elements exhibit an exponential concentration~\cite{thanasilp2022exponential}. Specifically, the  concentration of quantum kernel element $K^Q_{i,j} = k^Q(x_i,x_j)$ can be expressed by Chebyshev's inequality,
\begin{equation}
\mathrm{Pr} \left[ \left\vert K^Q_{i,j} - \mathbb{E}\left[ K^Q_{i,j}\right] \right\vert \geq \delta \right]\leq \frac{\mathrm{Var}\left[K^Q_{i,j}\right]}{\delta^2}
\end{equation}
for any $\delta > 0$. The quantum kernel element $K^Q_{i,j}$ arising from highly expressive quantum embedding displays an exponentially reduction in variance as the number of qubits increases. Consequently, an exponentially large number of quantum circuit executions is necessary to accurately approximate the quantum kernel matrix $K^Q$. This poses a significant challenge in the efficient implementation of QKM.

NQE addresses this challenge by constraining quantum embedding to ensure large distinguishability. NQE strategically limits the expressibility of the embedded quantum states, thereby enhancing the trainability of QML models. This improvement is achieved by exploiting the prior knowledge that a quantum embedding with large distinguishability can effectively approximate the true underlying function. 

\begin{figure}[t]
    \centering
\includegraphics[width=1.0\columnwidth]{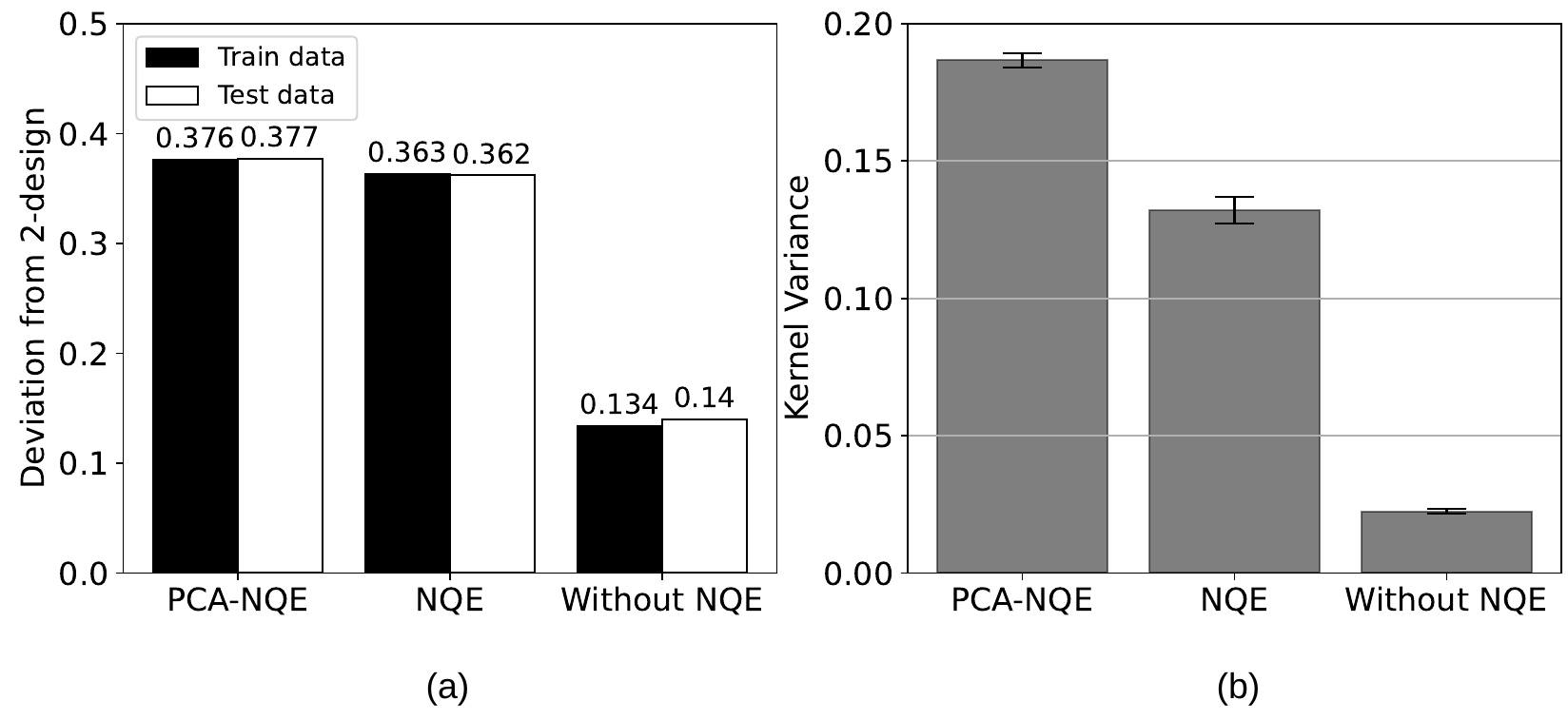}
    \caption{
(a) A comparative analysis of expressibility with and without NQE models. The deviation from unitary 2-design is depicted, where a smaller deviation indicates higher expressibility. The deviation is derived from 12,665 (2,115) MNIST training (test) data results. (b) A comparative analysis of the variance of quantum kernel elements with and without NQE models. The variance was computed from the off-diagonal elements of the quantum kernel matrix $K^Q$, constructed from 1,000 samples of MNIST datasets. The mean and one standard deviation from five independent iterations are shown. For both (a) and (b), NQE models are optimized using the ibmq\_toronto quantum device.
    } \label{fig:ExpTrain}
\end{figure}

Figure~\ref{fig:ExpTrain}(a) illustrates how the expressibility varies as we apply NQE models. Here, we investigated the Hilbert-Schmidt norm of the deviation from unitary 2-design~\cite{Sim_expressibility, holmes_2022}, as a measure of expressibility. More specifically, the deviation is given as, 
\begin{equation}
A = \int_{\textbf{Haar}} (\vert \psi \rangle \langle \psi \vert)^{\otimes2} d\psi -\int_{\mathcal{E}} (\vert \phi \rangle \langle \phi \vert)^{\otimes2} d\phi,
\end{equation}
where the first integral is taken over the Haar measure, and the second integral is taken over ensemble of data embedded quantum states, $\mathcal{E}$. We then define the deviation norm, $\epsilon = \sqrt{\text{Tr}(A^\dagger A)}$, where a small $\epsilon$ indicates a highly expressive quantum embedding and vice versa. In this experiment, we employed NQE and PCA-NQE, which had been previously optimized with ibmq\_toronto hardware as detailed in Section~\ref{sec:NQEvsFUE}. The value of  $\epsilon$ was numerically computed for ensembles of training and test datasets of classes \{0,1\} from MNIST data, utilizing 12,665 instances for training data and 2,115 for test data. We observe that for both training and test data, both NQE methods lead to reduction in the expressibility, consequently enhancing trainability.

Figure~\ref{fig:ExpTrain}(b) demonstrates how the variance of the quantum kernel elements varies as we apply NQE models. To determine the variance, we initially computed quantum kernel matrix $K^Q$ using PCA-NQE, NQE, and quantum embedding without NQE, following the procedures outlined in Section~\ref{sec:QKM}. The variance was calculated using the off-diagonal elements of $K^Q$. These experiments were repeated five times, each with different 1,000 samples of data.  The figure display the mean and one-standard deviation calculated from these results.
A significant increase in the variance of quantum kernel elements is observed with the use of NQE models, implying that the quantum kernel matrix $K^Q$ can be reliably approximated with fewer quantum circuit executions when NQE models are implemented. 
These results underscores the effectiveness of NQE in enhancing the trainability of QML models within both QNN and QKM frameworks.

\section{Conclusions and Discussion}
\label{sec:discussion}
In this study, we investigated the crucial role of quantum embedding, an essential step in applying quantum machine learning to classical data. In particular, we highlighted how quantum data separability, namely the distinguishability of quantum states representing different classes, determines the lower bound of training error and affects the noise resilience of quantum supervised learning algorithms. Motivated by these results, we introduced Neural Quantum Embedding (NQE), which utilizes the power of classical neural network and deep learning to enhance data separability in the quantum feature space, thereby pushing the limits of quantum supervised learning. Integrating classical neural networks with parameterized quantum circuits to construct ML models has been explored in various studies, as referenced in~\cite{broughton2020tensorflow,Mari2020transferlearningin,co-design,9951229,kim2023classical}. However, there remains ambiguity regarding why and how classical neural networks should be incorporated, aside from addressing the limited size of quantum circuits executable on NISQ computers or naively attempting to extend the success of deep learning to the domain of QML. Additionally, determining the optimal strategy for interfacing classical and quantum neural networks, particularly for transferring information from classical to quantum systems, remains an important open problem. This work bridges these gaps by linking quantum supervised learning with the theory of quantum state discrimination, and contributes to developing an effective approach for applying QML to classical data. Quantum supervised learning for data classification can be understood as a process of learning the optimal POVM for minimizing error in discriminating density matrices representing data samples across different classes. In this regard, the parameterized quantum circuit solely plays the role of identifying the optimal measurement, whereas the optimal training performance is dictated by how the data is encoded as the quantum
state. This optimal performance cannot be improved by any PTP map (e.g. quantum channels). However, classical neural networks can be utilized to learn the feature map from data samples, maximizing the distinguishability of states in the quantum feature space beyond the limits of the PTP maps for the given dataset. A quantum computer is also essential in this task because the objective function of the classical neural network is based on the fidelity of quantum states, which is conjectured to be hard to compute classically for the quantum feature maps of the interest. In this respect, NQE differs from existing classical-quantum neural networks in that it trains a classical neural network with an objective function computed by a quantum computer specifically to maximize the separability of quantum data, which is the optimal embedding strategy for classification tasks.

NQE is versatile in the sense that it can be integrated into all existing quantum data embedding methods, such as amplitude encoding, angle encoding, and Hamiltonian encoding. The training performance achieved by NQE is guaranteed to be at least as good as those that do not use it and rely on a fixed embedding function. This is because if the fixed embedding function happens to be the optimal one for the given data, NQE will learn to use it. Experimental results on IBM quantum hardware demonstrate that NQE significantly enhances quantum data separability, as quantified by increased trace distance between two ensembles of quantum states. Utilizing NQE led to a significant reduction in training loss and an improvement in accuracy and noise resilience in the MNIST data classification tasks. Notably, the experimental results achieved by NQE-enabled QML models outperformed the theoretical optimal of the conventional ZZ feature embedding that does not employ NQE.

Furthermore, we conducted numerical comparisons between NQE and three trainable unitary embedding circuits using both MNIST and Fashion-MNIST datasets. This study encompassed both noiseless and noisy simulations. The results demonstrate that NQE outperforms trainable unitary embeddings in terms of both training and classification accuracies across all scenarios.

A significant portion of current research in QML focuses on the trade-off between the expressibility and trainability of variational circuits within quantum neural networks. For a QML model to be effective, it must possess a high degree of expressibility, which ensures that it can approximate the desired solution with considerable accuracy. Concurrently, the model should be trainable, enabling optimization via a gradient descent algorithm or its variants. However, expressibility and trainability present a trade-off: high expressibility typically leads to reduced trainability ~\cite{holmes_2022, thanasilp2022exponential, 10.1038/s41467-018-07090-4}. This trade-off constitutes a significant challenge in advancing QML. To address this challenge, a strategic approach is to utilize problem-specific prior knowledge. For example, Refs.~\cite{larocca2022group, meyer2023exploiting} deliberately construct variational circuits with limited expressibility, yet ensures the inclusion of the desired solution, by harnessing data symmetry. However, such method is not universally applicable to general datasets that do not presents any symmetry or group structure. NQE offers an effective solution to this challenge by optimizing the quantum data embedding process. As elucidated in the Results section, a good approximation of the true underlying function can only be achieved with quantum embeddings that ensure high distinguishability of the data. By using this prior knowledge, NQE constrain the quantum embedding to those that allows large distinguishability between quantum states that represent the data. Consequently, the embedded quantum states from NQE are less expressive, resulting in an improvement in trainability.

The ultimate goal of machine learning is to construct a model that not only classifies the training data accurately (optimization) but also generalizes well to unseen data. In conventional machine learning, a trade-off typically exists between optimization and generalization~\cite{vapnik1999nature}. However, our experimental results indicate that the incorporation of NQE markedly enhances both optimization and generalization metrics. This improvement is evidenced by reduced training error, reduced test error, diminished local effective dimension, and a reduced generalization upper bound in quantum kernel method. Consequently, NQE presents a robust methodology for optimizing learning performance while preserving strong generalization capabilities.

Appendices~\ref{appendix:FMNIST} and~\ref{appendix:scalability} present additional experimental findings. The former exhibits noiseless and noisy simulation outcomes on the Fashion-MNIST dataset, illustrating the advantages of NQE with an alternative dataset. The latter delves into 8- and 12-qubit noiseless simulation results obtained from MNIST datasets. These results demonstrate that the benefits of NQE persist even for larger quantum systems. Overall, the supplementary findings consistently affirm that NQE surpasses traditional quantum data embedding methods.

Further research is necessary to explore the impact of the type or architecture of neural networks on the performance of NQE, and its optimization for specific types of target data. For instance, investigating the applicability of Recurrent Neural Networks (RNNs) for handling sequential data and Convolutional Neural Networks (CNNs) for image data remains an interesting avenue for future work. The incorporation of structure learning introduced in Ref.~\cite{incudini2022structure} with NQE is noteworthy as it can further improve the embedding. However, one must consider the trade-off between performance and the computational overhead introduced by the structure learning. As an alternative to NQE, enhancing quantum data separability can be achieved by implementing a probabilistic non-TP embedding~\cite{kwon2023feature}. Comparing this approach with NQE or exploring their combination for potential enhancements represents a valuable direction for future investigation.

\section*{Data Availability}

The data and software that support the findings of this study can be found in the following repository: \url{https://github.com/qDNA-yonsei/Neural-Quantum-Embedding}.

\section*{Acknowledgments}
This work was supported by Institute of Information \& communications Technology Planning \& evaluation (IITP) grant funded by the Korea government (No. 2019-0-00003, Research and Development of Core technologies for Programming, Running, Implementing and Validating of Fault-Tolerant Quantum Computing System), the Yonsei University Research Fund of 2023 (2023-22-0072), the National Research Foundation of Korea (Grant No. 2022M3E4A1074591, 2023M3K5A1094805), and the KIST Institutional Program (2E32241-23-010).

\appendix

\section{Experimental details}
\subsection{NQE structures and training}
\label{appendix:1}
The PCA-NQE method employs Principal Component Analysis (PCA) to extract $n$ features from the original data, where $n$ is the number of qubits used for data embedding. These features are then passed to a fully connected neural network with two hidden layers. In the case of four-qubit experiments, each hidden layer contains 12 nodes, whereas in the eight-qubit cases, each hidden layer contains 24 nodes. The neural network has $2n$ output nodes, corresponding to $2n$ numerical values used as quantum gate parameters for the embedding. The rectified linear unit (ReLU) function serves as the non-linear activation. In contrast, NQE (without PCA) utilizes a 2-dimensional convolutional neural networks (CNN) which takes original data as an input. After each convolutional layer, we used 2D max pooling to reduce the dimension of the data. The dimension of the nodes in each layer are 28 by 28, 14 by 14, 7 by 7, and $2n$, respectively.

The classical neural networks of the NQE models are optimized by minimizing the implicit loss function $l_{\mathrm{fid}}((x_i,y_i),(x_j,y_j))$ given in Eq.~(\ref{eq:loss}), where $i$ and $j$ are the indices of the randomly selected training data. For the four-qubit real device experiments, the NQE models were trained for 50 iterations using the stochastic gradient descent with a learning rate of 0.1 and a batch size of 10. The loss function was evaluated on ibmq\_toronto with the selection of four qubits based on the highest CNOT fidelities. For the ZZ feature embedding circuits, we configured the total number of layer ($L$) to 1 and applied two qubit gates only on the nearest neighboring qubits to avoid an excessive number of CNOT gates.

\subsection{Classification with QCNN}
\label{appendix:2}
In the noiseless simulation setting, we utilized QCNN circuits featuring a general SU(4) convolutional ansatz (refer to Fig. 2 (i) in Ref.~\cite{hur2022quantum}). The optimization of circuit parameters was performed over 1,000 iterations using the Nesterov momentum algorithm with a learning rate of 0.01 and a batch size of 128. Each simulation was repeated five times with random parameter initialization.

For experiments on IBM quantum hardware, QCNN circuits were configured with a basic convolutional ansatz comprising two $R_y(\theta)$ gates, where $R_i(\theta)$ represent a single-qubit rotation around the $i$-axis of the Bloch sphere by an angle $\theta$, and a CNOT gate (refer to Fig. 2 (a) in Ref.~\cite{hur2022quantum}). To minimize circuit depth, pooling gates were omitted. Moreover, the QCNN architecture was designed to allow only nearest-neighbor qubit interactions, eliminating the need for qubit swapping.

The training on quantum hardware consisted of 50 optimization iterations, using Nesterov momentum gradient descent with a learning rate of 0.1 and a batch size of 10. Experiments were conducted on three distinct quantum devices: ibmq\_jakarta, ibmq\_toronto, and ibmq\_perth.

Performance evaluation was carried out by assessing the classification accuracy of the trained QCNN models on a separate test set comprising 500 data points. This assessment was executed across three different quantum devices: ibmq\_lagos, ibmq\_kolkata, and ibmq\_jakarta. The results presented were obtained from 1,024 executions of quantum circuits.

\subsection{Effective Dimension}
\label{appendix:3}

\begin{figure}[t]
    \centering
    \includegraphics[width=0.7\columnwidth]{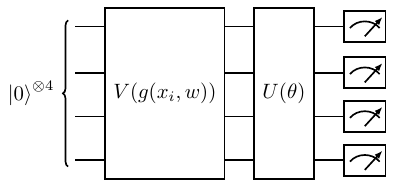}
    \caption{Quantum circuit used for evaluating the local effective dimension.
    The feature map, denoted by $V(g(x_i,w))$, acts on the initial state $\ket{0}^{\otimes 4}$ to encode the input vector $x_i$. 
    Subsequently, a parameterized unitary transformation $U(\theta)$ is applied to evolve the state, with the parameters $\theta$ chosen to minimize a specific loss function.  
    }  \label{fig:LED}
\end{figure}

\begin{figure}[t]
    \centering
    \subfloat{
        \includegraphics[width=0.80\linewidth]{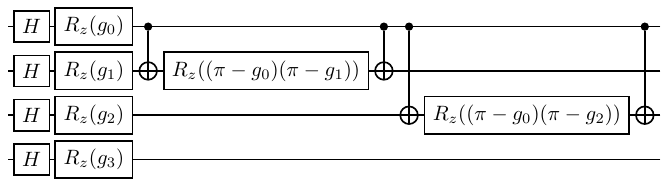}
        } \\
   
    \subfloat{
        \includegraphics[width=\linewidth]{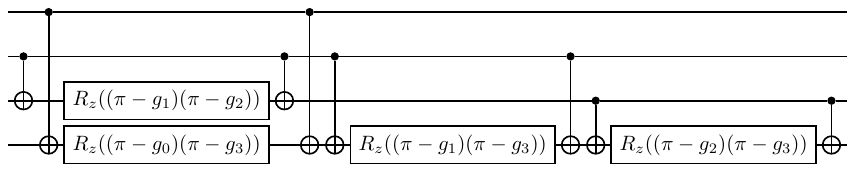}
    }
    \caption{Quantum circuit diagram for the trainable ZZ feature map $V(g(x_i,w))$. The top and bottom segments collectively represent a single quantum circuit, split into two parts due to limited horizontal space. It uses mapping functions $g_j \in g(x_i,w)$ to encode the input vector $x_i$.
    }  \label{fig:feature_map}
\end{figure}

The analysis of the effective dimension utilizes the QNN architecture depicted in Fig.~\ref{fig:LED}. The ZZ feature map, as explained in Eq.~\ref{eq:qembed} and depicted in Fig.~\ref{fig:feature_map}, is used for mapping classical data to quantum states. When NQE is not used, $g_j = x_{ij}$, which is the $j$th component of the input vector $x_i$, and the quantum embedding becomes equivalent to Eq.~(\ref{eq:qembed}) with $L=1$. When NQE is turned on, $g_j = g_j(x_i,w)$, which is the $j$th component of the output vector generated by a three-layer neural network. This output vector can be expressed as $g(x_i,w) = \sigma ( w^{(1)} { \sigma ( w^{(0)} x_i+b^{(0)} ) }+b^{(1)} )$. In this equation, $w^{(0)}\in\mathbb{R}^{12\times 4}$, $w^{(1)}\in\mathbb{R}^{4\times 12}$, $b^{(0)}\in\mathbb{R}^{12}$ and $b^{(1)}\in\mathbb{R}^{4}$ are the trainable parameters of the network, and $\sigma$ stands for the ReLU activation function.

\begin{figure}[t]
    \centering
    \includegraphics[width=1.0\columnwidth]{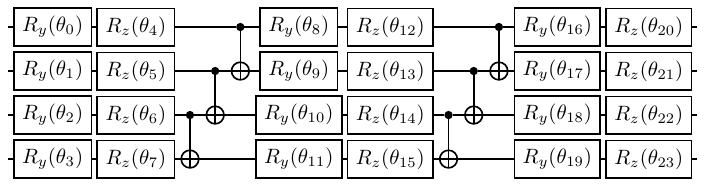}
    \caption{Parameterized quantum circuit layout for the QNN, which is represented as $U(\theta)$ in Fig.~\ref{fig:LED}.
    }  \label{fig:variational_model}
\end{figure}

The parameterized unitary operator of the QNN used in this analysis, represented as $U(\theta)$ Fig.~\ref{fig:LED}, is shown in Fig.~\ref{fig:variational_model}. This circuit design is attractive for several reasons. First, it offers a high degree of expressibility and entangling capability while maintaining a relatively small number of gates and parameters~\cite{Sim_expressibility}. In addition, this design is hardware-efficient~\cite{kandala_hardware-efficient_2017,cerezo2020variational}, as it relies solely on single-qubit rotations and CNOT operations between adjacent qubits.

The analysis encompassed ten artificial binary datasets, each containing 400 samples. These datasets, characterized by four features and four clusters per class, were generated through the \texttt{make\_classification} function from the Scikit-Learn library~\cite{scikit-learn}. 
The NQE model was trained for 100 iterations using the Adam optimizer~\cite{kingma2017adam} with a batch size of 25.

The local effective dimension~\cite{abbas2021_b} was determined using the \texttt{get\_effective\_dimension} function from the \texttt{LocalEffectiveDimension} class in Qiskit~\cite{qiskit2024}.

\section{Relation between linear and MSE loss}
The main text focuses on a maximizing the trace distance, which sets the optimal lower bound for the linear loss. However, many conventional quantum machine learning (QML) routines employ mean squared error (MSE) loss. Here,  we provide some relationship between linear loss and MSE loss. Consider the data $\{x_i, y_i\}_{i=1}^{N}$ and the vectors $Y=(y_1, y_2\ldots  y_N)$ and $f(X)= (f(x_1),f(x_2) \ldots f(x_N))$. The linear and MSE loss are expressed as, 
\begin{align}
L_{\mathrm{linear}} &= \vert\vert Y - f(X) \vert\vert_1 \\ L_{\mathrm{MSE}} &= \vert\vert Y - f(X) \vert\vert_{2}^2
\end{align}
By the vector norm inequalities, we can both upper bound and lower bound the MSE loss with linear loss,
\begin{align}
\frac{1}{N}L_{\mathrm{linear}}^2 \leq L_{\mathrm{MSE}} &\leq L_{\mathrm{linear}}^2.
\end{align}
Reducing the lower bound of empirical linear loss by maximizing trace distance, reduces both the upper and lower bound of the empirical MSE loss. Hence, we can expect neural quantum embedding (NQE) to work favorably for MSE loss as well.

\section{Relation between implicit loss function and trace distance}
\label{Appendix:C}
During the NQE training, we optimize the classical neural network to maximize the trace distance between two data embedded ensembles. Although using trace distance directly as a loss function is ideal, we utilized an implicit loss function due to the computational hardness of the trace distance calculation. The implicit loss function is delineated in equation~\ref{eq:loss}. 

When $y_i = y_j$, the loss function directs NQE to maximize the fidelity between $\vert x_i \rangle$ and $\vert x_j \rangle$ as much as possible. Due to the contractive property of the trace distance, $D(\rho^-, \rho^+) \leq D(\vert \psi^- \rangle, \vert \psi^+ \rangle) $, where, $\vert \psi^{-} \rangle, \vert \psi^+\rangle$ are purification of $\rho^-, \rho^+$, respectively. The equality holds when the two data ensembles are pure states. The purity of $\rho^{\pm} $ is 
$$\text{Tr}\left(\left(\rho^{\pm}\right)^2\right) = \frac{1}{\left(N^{\pm}\right)^2} \sum_{i,j=1}^{N^\pm} \left\vert \braket{x_i^\pm}{x_j^\pm} \right\vert^2. $$ 
Therefore, maximizing the fidelity when $y_i = y_j$ increases the purity of $\rho^{\pm}$, allowing the trace distance to achieve its upper bound. 

Conversely, when $y_i \neq y_j$, the loss function directs NQE to minimize the fidelity between $\vert x_i \rangle$ and $\vert x_j \rangle$ as much as possible. For simplicity, let’s consider a balanced set of data $N^+=N^-=N$. Due to strong convexity of the trace distance~\cite{wilde_2013},
\begin{equation}
D\left(\frac{1}{N}\sum_{i=1}^N \vert x_i^- \rangle \langle x_i^- \vert, \frac{1}{N}\sum_{i=1}^N \vert x_i^+ \rangle \langle x_i^+ \vert \right) \leq  \frac{1}{N} \sum_{i=1}^N \sqrt{1-\left\vert \braket{x_i^+}{x_i^-} \right\vert^2}
\end{equation}
Hence, minimizing the fidelity when $y_i \neq y_j$ increases the upper bound of the trace distance. Therefore, it is evident that minimizing the implicit loss function contributes positively to maximizing the trace distance.

\section{Simulation Results on an Additional Dataset}
\label{appendix:FMNIST}
In Section~\ref{sec:NQEvsFUE} of the main text, we presented how utilizing PCA-NQE and NQE improves QCNN performance when classifying MNIST datasets, by demonstrating trace distance history (Figure~\ref{fig:qcnn_demonstration}(a)), QCNN loss history and classification accuracies (Figure~\ref{fig:qcnn_demonstration}(b) and (c)). In Section~\ref{sec:QKM} and Section~\ref{sec:ExpTrain}, we illustrated how employing NQE models can improve generalization performances and trainability. 

\begin{figure*}[ht]
    \centering
        \includegraphics[width=0.96\textwidth]{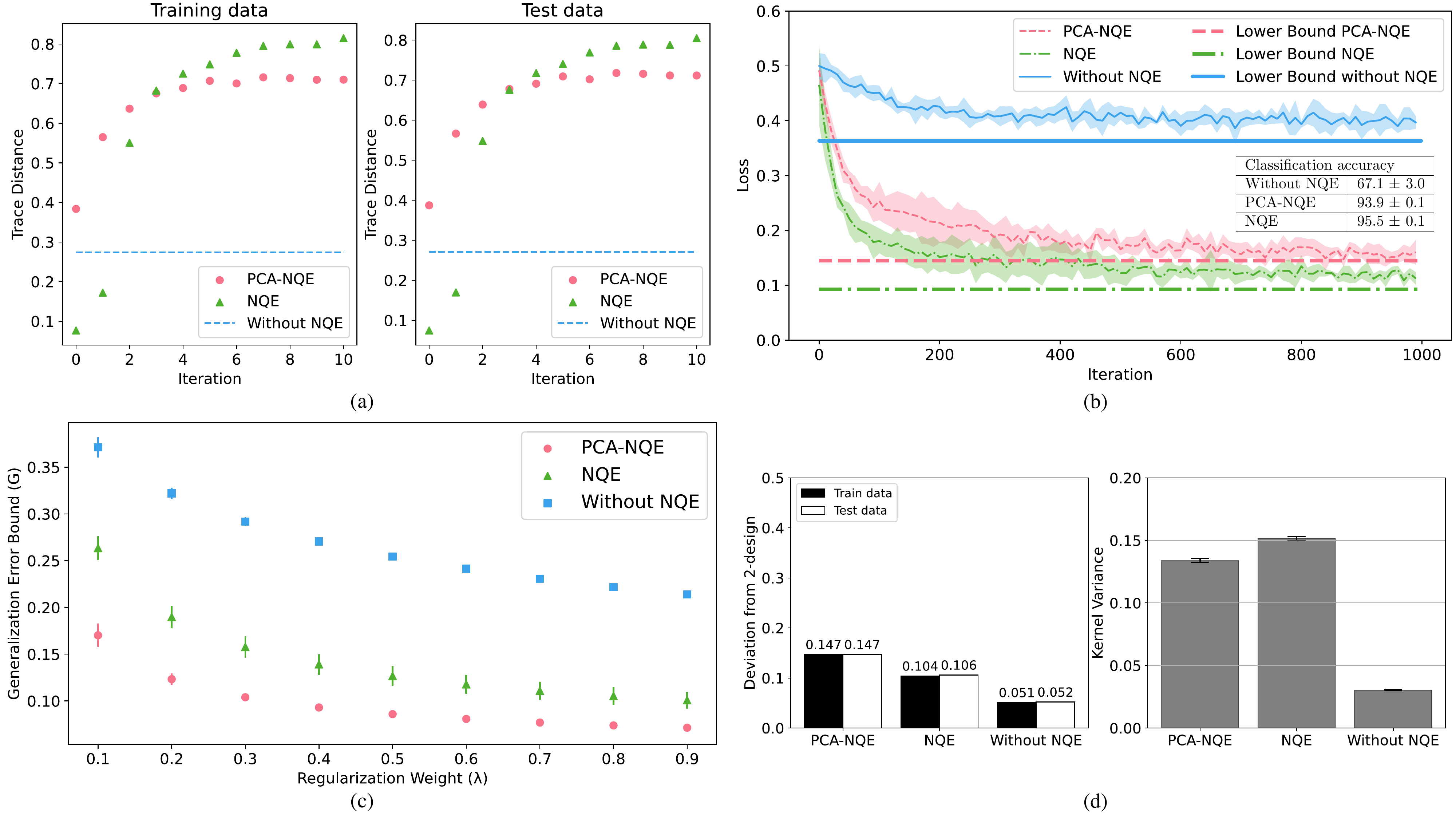}
        \caption{Results for the additional experiments with Fashion-MNIST datasets using noiseless simulations. (a) Plot depicting the evolution of the trace distance between two ensembles of quantum states embedded by the NQE models during training, compared to the trace distance from conventional quantum embedding without NQE. (b) QCNN simulation results. The blue solid, red dashed, and green dash-dotted lines represent the mean training loss histories for conventional ZZ feature embedding, PCA-NQE, and NQE, respectively. The shaded regions in the figure represent one standard deviation from the mean. These values are acquired from five repetitions of each QCNN training with random initialization of parameters. The thicker versions of these lines indicate the theoretical lower bounds for each method.  (c) A comparative analysis of the generalization error bound $G$ with varying regularization weights $\lambda$. This plot illustrates the performance enhancement---lower generalization error bound---when employing NQE (green triangles) and PCA-NQE (red circles) over conventional methods without NQE (blue squares) in quantum kernel methods. The error bound $G$ was determined based on five independent numerical simulations, presenting both the mean and one standard deviation of $G$. (d) Left: A comparative analysis of expressibility with and without NQE models. The deviation from unitary 2-design is depicted, where a smaller deviation indicates higher expressibility. The deviation is derived from 12,000 (2,000) Fashion-MNIST training (test) data results. Right: A comparative analysis of the variance of quantum kernel elements with and without NQE models. The variance was computed from the off-diagonal elements of the quantum kernel matrix $K^Q$, constructed from 1,000 samples of Fashion-MNIST datasets. The mean and one standard deviation from five independent iterations are shown.
        }
        \label{fig:qcnn_noiseless_fashion}
\end{figure*}

\begin{figure*}[ht]
    \centering
        \includegraphics[width=0.96\textwidth]{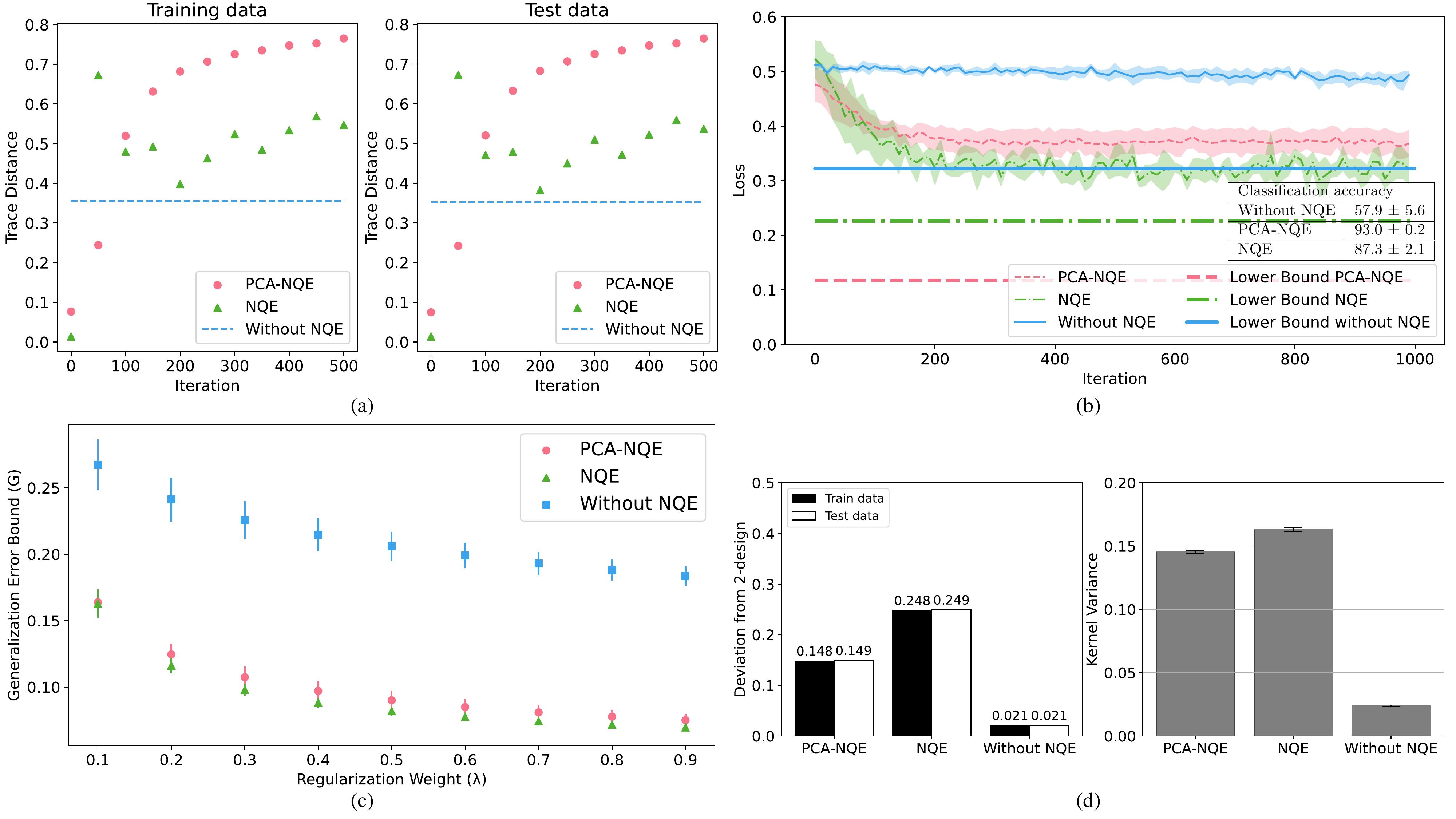}
        \caption{Results for the additional experiments with Fashion-MNIST datasets using noisy simulations replicating the characteristics of the ibmq\_vigo quantum computer. Descriptions for (a), (b), (c), and (d) are identical to those provided in the caption of Fig.~\ref{fig:qcnn_noiseless_fashion}.}
        \label{fig:qcnn_noisy_fashion}
\end{figure*}

In this section, we present additional experiments tested on classes $\{\text{T-shirt/Top},\text{Trouser}\}$ of Fashion-MNIST dataset~\cite{xiao2017/online}. Figure~\ref{fig:qcnn_noiseless_fashion} and~\ref{fig:qcnn_noisy_fashion} display results from noiseless and noisy simulations, respectively. In both figures, (a) depicts the trace distance history with and without NQE models, (b) presents QCNN loss history and classification accuracy with and without NQE models, (c) presents the upper bound of generalization error in QKM with and without NQE models and (d) presents expressibility and trainability with and without NQE models. In alignment with the results in the main text, additional experiments with Fashion-MNIST datasets indicate that employing NQE effectively improves training error, classification accuracy, generalization performance, and trainability.

The methodology for these experiments mirrors that of the experiments with MNIST datasets, which are detailed in Appendix~\ref{appendix:1}, and Sections ~\ref{sec:QKM} and ~\ref{sec:ExpTrain}. Due to the constraints in accessing IBM quantum hardware, we adapted our approach for the noisy experiments. Instead of direct hardware utilization, we employed a simulation environment using the IBM FakeVigo device. This simulator mimics the essential characteristics of the ibmq\_vigo device, including its basis gates, qubit connectivity, qubit relaxation (T$_1$) and dephasing (T$_2$) times, and readout error rates.

\section{Simulation Results with Larger Quantum Circuits}
\label{appendix:scalability}
\begin{figure*}[ht]
    \centering
        \includegraphics[width=0.96\textwidth]{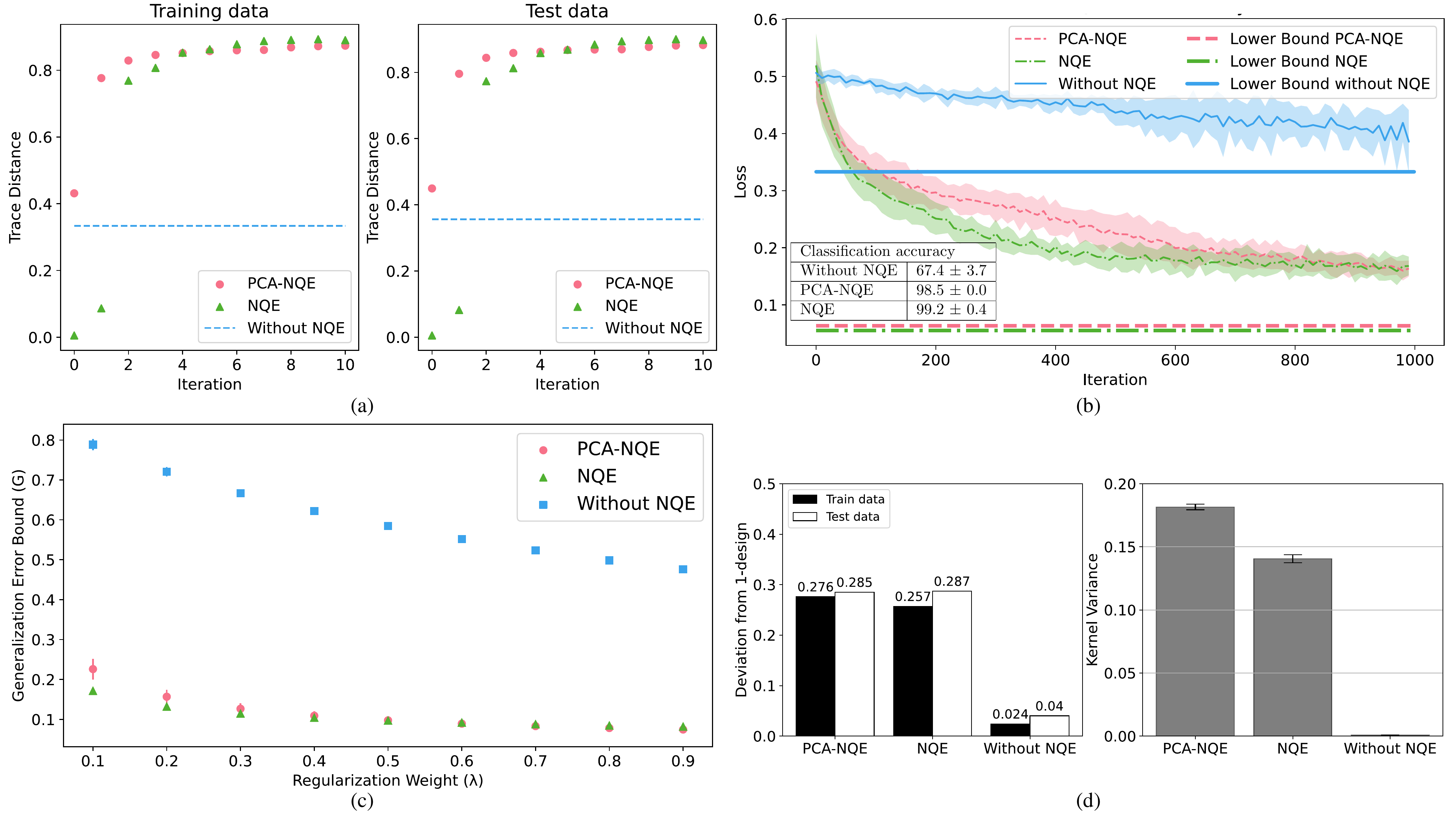}
        \caption{Results for the additional experiments with 8 qubit noiseless simulations. Descriptions for (a), (b), and (c) are identical to those provided in the caption of Fig.~\ref{fig:qcnn_noiseless_fashion}. (d) Left: A comparative analysis of expressibility with and without NQE models. The deviation from unitary 1-design is depicted, where a smaller deviation indicates higher expressibility. The deviation is derived from 12,665 (2,115) MNIST training (test) data results. Right: A comparative analysis of the variance of quantum kernel elements with and without NQE models. The variance was computed from the off-diagonal elements of the quantum kernel matrix $K^Q$, constructed from 1,000 samples of MNIST datasets. The mean and one standard deviation from five independent iterations are shown.}
        \label{fig:scalability 8 qubits}
\end{figure*}

\begin{figure*}[ht]
    \centering
        \includegraphics[width=0.96\textwidth]{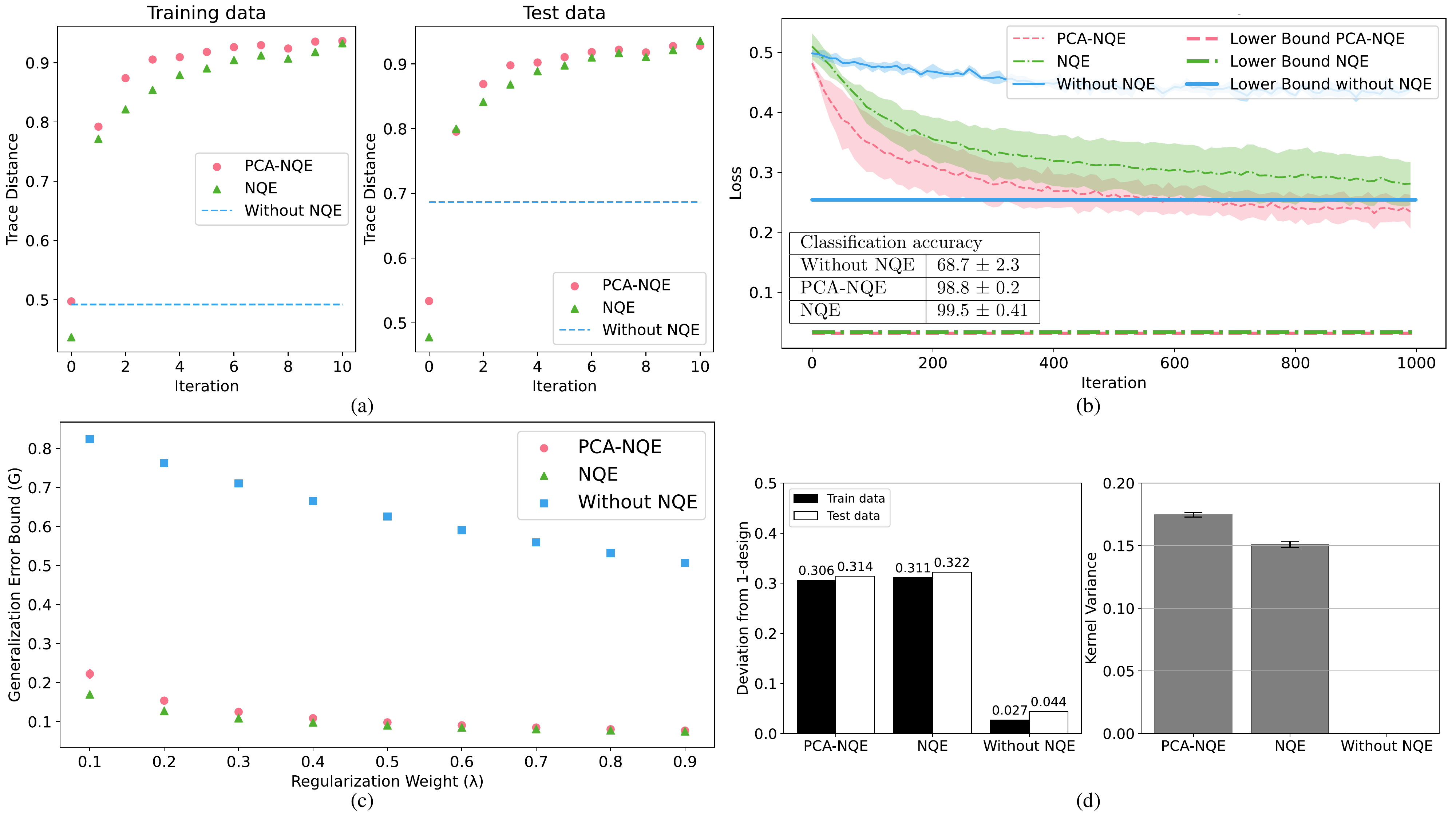}
        \caption{Results for the additional experiments with 12 qubit noiseless simulations. Descriptions for (a), (b), (c), and (d) are identical to those provided in the caption of Fig.~\ref{fig:scalability 8 qubits}.}
        \label{fig:scalability 12 qubits}
\end{figure*}

In Section~\ref{sec:NQEvsFUE} of the main text, we presented how utilizing NQE improves QCNN performances by demonstrating four-qubits experiments on both numerical simulation and IBM quantum hardware experiments. In this section, we illustrate the effectiveness of NQE methods in larger quantum systems, specifically those comprising eight and twelve qubits. Consistent with the main text, we compare effectiveness of NQE methods on trace distance history (Figure~\ref{fig:scalability 8 qubits},~\ref{fig:scalability 12 qubits}(a)), QCNN loss histories and classification accuracies (Figure~\ref{fig:scalability 8 qubits},~\ref{fig:scalability 12 qubits}(b)), upper bound of generalization in QKM (Figure~\ref{fig:scalability 8 qubits},~\ref{fig:scalability 12 qubits}(c)) and expressibility and trainability (Figure~\ref{fig:scalability 8 qubits},~\ref{fig:scalability 12 qubits}(d)) in 8 and 12 qubit setups. Due to limited computational resources and access to IBM quantum hardware the experiments are only conducted with numerical simulation. The experimental methods are identical to the ones of Section~\ref{sec:NQEvsFUE} of the main text, except the expressibility is computed by deviation from 1-design (instead of 2-design) due to limited computational resources. 

The experimental results further validates that NQE methods are effective at enhancing QML algorithms on larger quantum systems. Application of NQE yielded improvements in training loss, classification accuracy, generalization upper bounds, and trainability metrics. Unlike in four-qubit experiments, training losses did not reach their theoretical minima. This indicates that QCNN circuits did not accurately approximate the optimal Helstrom measure. Such behavior is expected as QCNN circuits are inexpressive due to its parameter-sharing and nearest-neighbor variational ansatz constraints. Nonetheless, NQE application significantly enhanced training loss, underscoring its utility in advancing QML algorithm performance. Additionally, in scenarios where NQE are not employed, there is a notable decline in trainability with 8 and 12 qubits (Fig.~\ref{fig:scalability 8 qubits},~\ref{fig:scalability 12 qubits} (d)), as opposed to those with 4 qubits (Fig.~\ref{fig:ExpTrain}). This reduction in trainability is evidenced by increased expressibility and kernel variance. Conversely, in systems utilizing NQE models, both expressibility and kernel variance maintain consistent levels, demonstrating enhanced trainability even in larger-scale quantum circuits.

\section{Additional Analysis on Expressibility and Trainability}
\label{appendix:expressibility}
Let $N$ be the number of samples drawn from a data distribution $\mathcal{D}$, and let $d=\mathrm{dim}(\mathrm{span}\lbrace |x_i\rangle\rbrace_{i=1,\ldots,N})\le N$ represents the dimension of the subspace spanned by the quantum state representation of training data, determined by quantum embedding~\cite{huang_power_2021}. The dimension $d$ serves as an indicator of the expressibility of the given quantum embedding and the complexity of the machine learning (ML) model necessary for learning from the quantum-embedded data. This relationship can be rigorously demonstrated in the context of the quantum kernel method (QKM) as follows. The QKM can be employed to learn a quantum model deﬁned by $f(x) = \Tr(OU\rho(x)U^{\dagger})$, where $\rho(x)$ is the density matrix representation of the data encoded as a quantum state, from $N$ samples drawn from a data distribution $\mathcal{D}$. The expected risk (i.e. prediction error) of the prediction model $h(x)$ constructed from the learning procedure is bounded as
\begin{equation} \label{eq:prediction_error_bound}
    \mathbb{E}_{x \in \mathcal{D}} \vert h(x) - f(x) \vert \leq c \sqrt{\frac{\min(d, \Tr(O^2))}{N}},
\end{equation}
where $c>0$ is a constant. The quantity of interest here is $d$, because it is affected by NQE and $\Tr(O^2)$ grows exponentially with the number of qubits in many cases (e.g. Pauli observables). This equation implies that the hardness of the learning problem depends on the quantum embedding that represents the set of $N$ training data.

\begin{table}[h]
    \resizebox{0.8\columnwidth}{!}{
    \begin{tabular}{cc|cccccc|}
\cline{3-8}
                                               &            & \multicolumn{6}{c|}{$d$}                                                                                                                                        \\ \hline
\multicolumn{2}{|c|}{Dataset}                               & \multicolumn{1}{c|}{\multirow{2}{*}{Untrained}} & \multicolumn{5}{c|}{\begin{tabular}[c]{@{}c@{}}NQE\\ (5 runs)\end{tabular}}                                 \\ \cline{1-2} \cline{4-8} 
\multicolumn{1}{|c|}{Classes}                  & Input Size & \multicolumn{1}{c|}{}                           & \multicolumn{1}{c|}{1}  & \multicolumn{1}{c|}{2}  & \multicolumn{1}{c|}{3}  & \multicolumn{1}{c|}{4}  & 5   \\ \hline
\multicolumn{1}{|c|}{\multirow{2}{*}{0 and 1}} & 4          & \multicolumn{1}{c|}{175}                        & \multicolumn{1}{c|}{15} & \multicolumn{1}{c|}{14} & \multicolumn{1}{c|}{19} & \multicolumn{1}{c|}{15} & 13  \\
\multicolumn{1}{|c|}{}                         & 8          & \multicolumn{1}{c|}{800}                        & \multicolumn{1}{c|}{14} & \multicolumn{1}{c|}{16} & \multicolumn{1}{c|}{17} & \multicolumn{1}{c|}{20} & 20  \\ \hline
\multicolumn{1}{|c|}{\multirow{2}{*}{3 and 8}} & 4          & \multicolumn{1}{c|}{175}                        & \multicolumn{1}{c|}{38} & \multicolumn{1}{c|}{38} & \multicolumn{1}{c|}{40} & \multicolumn{1}{c|}{31} & 27  \\
\multicolumn{1}{|c|}{}                         & 8          & \multicolumn{1}{c|}{800}                        & \multicolumn{1}{c|}{55} & \multicolumn{1}{c|}{54} & \multicolumn{1}{c|}{87} & \multicolumn{1}{c|}{50} & 127 \\ \hline
\end{tabular}
    }
    \caption{ \label{tab:kernel_rank}The rank of the quantum kernel matrix ($d$) evaluated with and without NQE. Four different binary datasets, each containing 800 samples, were generated from two pairs of MNIST classes. For each pair of datasets, two PCA configurations were used to reduce the number of features to 4 and 8, respectively.}
\end{table}

We conducted numerical experiments to assess the effectiveness of NQE in reducing the dimension $d$. To evaluate this, we employed a binary classification task involving the discrimination of digits 0 and 1 or 3 and 8 from the MNIST dataset. We computed the rank of the quantum kernel matrix both with and without NQE, and the dimension $d$ was determined as $d=\mathrm{rank}(K^Q)$. The results, presented in Table~\ref{tab:kernel_rank}, clearly demonstrate that NQE effectively reduces $d$. As $d$ represents the effective dimension of the quantum training data used for model training, its reduction indicates that simpler ML models with NQE can achieve comparable performance to more complex models applied to the original data without NQE. The findings suggest that, by using NQE, we can constrain quantum embedding to those that allow for large data separability. This reduction in the expressibility of quantum embedding, conversely, improves the trainability of the model.









\clearpage
\begin{thebibliography}{10}

\bibitem{Rebentrost_2018}
Patrick Rebentrost, Adrian Steffens, Iman Marvian, and Seth Lloyd.
\newblock Quantum singular-value decomposition of nonsparse low-rank matrices.
\newblock {\em Physical Review A}, 97(1), 2018.

\bibitem{PhysRevLett.113.130503_QSVM}
Patrick Rebentrost, Masoud Mohseni, and Seth Lloyd.
\newblock Quantum support vector machine for big data classification.
\newblock {\em Phys. Rev. Lett.}, 113:130503, Sep 2014.

\bibitem{lloyd2013quantum}
Seth Lloyd, Masoud Mohseni, and Patrick Rebentrost.
\newblock Quantum algorithms for supervised and unsupervised machine learning.
\newblock {\em arXiv preprint arXiv:1307.0411}, 2013.

\bibitem{QML-Biamonte}
Jacob Biamonte, Peter Wittek, Nicola Pancotti, Patrick Rebentrost, Nathan
  Wiebe, and Seth Lloyd.
\newblock Quantum machine learning.
\newblock {\em Nature}, 549:195 EP --, 09 2017.

\bibitem{10.1038/s43588-022-00311-3}
M.~Cerezo, Guillaume Verdon, Hsin-Yuan Huang, Lukasz Cincio, and Patrick~J.
  Coles.
\newblock {Challenges and opportunities in quantum machine learning}.
\newblock {\em Nature Computational Science}, pages 1--10, 2022.

\bibitem{aaronson2011computational}
Scott Aaronson and Alex Arkhipov.
\newblock The computational complexity of linear optics.
\newblock In {\em Proceedings of the forty-third annual ACM symposium on Theory
  of computing}, pages 333--342, 2011.

\bibitem{Lund2017npjQI}
A.~P. Lund, Michael~J. Bremner, and T.~C. Ralph.
\newblock Quantum sampling problems, bosonsampling and quantum supremacy.
\newblock {\em npj Quantum Information}, 3(1):15, 2017.

\bibitem{HarrowMondanaro2017QCS}
Aram~W. Harrow and Ashley Montanaro.
\newblock Quantum computational supremacy.
\newblock {\em Nature}, 549(7671):203--209, 2017.

\bibitem{arute2019quantum}
Frank Arute, Kunal Arya, Ryan Babbush, Dave Bacon, Joseph~C Bardin, Rami
  Barends, Rupak Biswas, Sergio Boixo, Fernando~GSL Brandao, David~A Buell,
  et~al.
\newblock Quantum supremacy using a programmable superconducting processor.
\newblock {\em Nature}, 574(7779):505--510, 2019.

\bibitem{zhong2020quantum}
Han-Sen Zhong, Hui Wang, Yu-Hao Deng, Ming-Cheng Chen, Li-Chao Peng, Yi-Han
  Luo, Jian Qin, Dian Wu, Xing Ding, Yi~Hu, et~al.
\newblock Quantum computational advantage using photons.
\newblock {\em Science}, 370(6523):1460--1463, 2020.

\bibitem{madsen2022quantum}
Lars~S Madsen, Fabian Laudenbach, Mohsen~Falamarzi Askarani, Fabien Rortais,
  Trevor Vincent, Jacob~FF Bulmer, Filippo~M Miatto, Leonhard Neuhaus, Lukas~G
  Helt, Matthew~J Collins, et~al.
\newblock Quantum computational advantage with a programmable photonic
  processor.
\newblock {\em Nature}, 606(7912):75--81, 2022.

\bibitem{Preskill2018quantumcomputingin}
John Preskill.
\newblock Quantum {C}omputing in the {NISQ} era and beyond.
\newblock {\em {Quantum}}, 2:79, August 2018.

\bibitem{cong_quantum_2019}
Iris Cong, Soonwon Choi, and Mikhail~D. Lukin.
\newblock Quantum convolutional neural networks.
\newblock {\em Nature Physics}, 15(12):1273--1278, December 2019.

\bibitem{benedetti_parameterized_2019}
Marcello Benedetti, Erika Lloyd, Stefan Sack, and Mattia Fiorentini.
\newblock Parameterized quantum circuits as machine learning models.
\newblock {\em Quantum Science and Technology}, 2019.

\bibitem{abbas2021_a}
Amira Abbas, David Sutter, Christa Zoufal, Aurelien Lucchi, Alessio Figalli,
  and Stefan Woerner.
\newblock {The power of quantum neural networks}.
\newblock {\em Nature Computational Science}, 1(6):403--409, 2021.

\bibitem{cerezo2020variational}
M.~Cerezo, Andrew Arrasmith, Ryan Babbush, Simon~C. Benjamin, Suguru Endo,
  Keisuke Fujii, Jarrod~R. {McClean}, Kosuke Mitarai, Xiao Yuan, Lukasz Cincio,
  and Patrick~J. Coles.
\newblock Variational quantum algorithms.
\newblock {\em Nature Reviews Physics}, 3(9):625--644, 2021.

\bibitem{Havlicek2019}
Vojtech Havl{\'i}cek, Antonio~D. C{\'o}rcoles, Kristan Temme, Aram~W. Harrow,
  Abhinav Kandala, Jerry~M. Chow, and Jay~M. Gambetta.
\newblock Supervised learning with quantum-enhanced feature spaces.
\newblock {\em Nature}, 567(7747):209--212, 2019.

\bibitem{RigorousRobustQSpeedUp}
Yunchao Liu, Srinivasan Arunachalam, and Kristan Temme.
\newblock A rigorous and robust quantum speed-up in supervised machine
  learning.
\newblock {\em Nature Physics}, 17(9):1013--1017, 2021.

\bibitem{schuld2021effect}
Maria Schuld, Ryan Sweke, and Johannes~Jakob Meyer.
\newblock Effect of data encoding on the expressive power of variational
  quantum-machine-learning models.
\newblock {\em Physical Review A}, 103(3):032430, 2021.

\bibitem{caro2021encoding}
Matthias~C Caro, Elies Gil-Fuster, Johannes~Jakob Meyer, Jens Eisert, and Ryan
  Sweke.
\newblock Encoding-dependent generalization bounds for parametrized quantum
  circuits.
\newblock {\em Quantum}, 5:582, 2021.

\bibitem{thanasilp2022exponential}
Supanut Thanasilp, Samson Wang, Marco Cerezo, and Zo{\"e} Holmes.
\newblock Exponential concentration and untrainability in quantum kernel
  methods.
\newblock {\em arXiv preprint arXiv:2208.11060}, 2022.

\bibitem{schuld2019quantum}
Maria Schuld and Nathan Killoran.
\newblock Quantum machine learning in feature hilbert spaces.
\newblock {\em Physical review letters}, 122(4):040504, 2019.

\bibitem{Nielsen:2011:QCQ:1972505}
Michael~A. Nielsen and Isaac~L. Chuang.
\newblock {\em Quantum Computation and Quantum Information: 10th Anniversary
  Edition}.
\newblock Cambridge University Press, New York, NY, USA, 10th edition, 2011.

\bibitem{wilde_2013}
Mark~M. Wilde.
\newblock {\em Quantum Information Theory}.
\newblock Cambridge University Press, 2013.

\bibitem{lloyd2020quantum}
Seth Lloyd, Maria Schuld, Aroosa Ijaz, Josh Izaac, and Nathan Killoran.
\newblock Quantum embeddings for machine learning.
\newblock {\em arXiv preprint arXiv:2001.03622}, 2020.

\bibitem{PhysRevA.106.042431}
Thomas Hubregtsen, David Wierichs, Elies Gil-Fuster, Peter-Jan H.~S. Derks,
  Paul~K. Faehrmann, and Johannes~Jakob Meyer.
\newblock Training quantum embedding kernels on near-term quantum computers.
\newblock {\em Phys. Rev. A}, 106:042431, Oct 2022.

\bibitem{10.1038/s41467-018-07090-4}
Jarrod~R. McClean, Sergio Boixo, Vadim~N. Smelyanskiy, Ryan Babbush, and
  Hartmut Neven.
\newblock {Barren plateaus in quantum neural network training landscapes}.
\newblock {\em Nature Communications}, 9(1):4812, 2018.

\bibitem{Bae_2015}
Joonwoo Bae and Leong-Chuan Kwek.
\newblock Quantum state discrimination and its applications.
\newblock {\em Journal of Physics A: Mathematical and Theoretical},
  48(8):083001, jan 2015.

\bibitem{e23050625}
Katarzyna Siudzińska, Sagnik Chakraborty, and Dariusz Chruściński.
\newblock Interpolating between positive and completely positive maps: A new
  hierarchy of entangled states.
\newblock {\em Entropy}, 23(5), 2021.

\bibitem{glick2021covariant}
Jennifer~R Glick, Tanvi~P Gujarati, Antonio~D Corcoles, Youngseok Kim, Abhinav
  Kandala, Jay~M Gambetta, and Kristan Temme.
\newblock Covariant quantum kernels for data with group structure.
\newblock {\em arXiv preprint arXiv:2105.03406}, 2021.

\bibitem{PhysRevLett.87.167902}
Harry Buhrman, Richard Cleve, John Watrous, and Ronald de~Wolf.
\newblock Quantum fingerprinting.
\newblock {\em Phys. Rev. Lett.}, 87:167902, Sep 2001.

\bibitem{huang_power_2021}
Hsin-Yuan Huang, Michael Broughton, Masoud Mohseni, Ryan Babbush, Sergio Boixo,
  Hartmut Neven, and Jarrod~R. McClean.
\newblock Power of data in quantum machine learning.
\newblock {\em Nature Communications}, 12(1):2631, 2021.

\bibitem{VQASVM}
Siheon Park, Daniel~K. Park, and June-Koo~Kevin Rhee.
\newblock Variational quantum approximate support vector machine with inference
  transfer.
\newblock {\em Scientific Reports}, 13(1):3288, 2023.

\bibitem{doi:10.1021/acs.jcim.1c00166}
Kushal Batra, Kimberley~M. Zorn, Daniel~H. Foil, Eni Minerali, Victor~O.
  Gawriljuk, Thomas~R. Lane, and Sean Ekins.
\newblock Quantum machine learning algorithms for drug discovery applications.
\newblock {\em Journal of Chemical Information and Modeling}, 61(6):2641--2647,
  2021.
\newblock PMID: 34032436.

\bibitem{Mensa_2023}
Stefano Mensa, Emre Sahin, Francesco Tacchino, Panagiotis~Kl Barkoutsos, and
  Ivano Tavernelli.
\newblock Quantum machine learning framework for virtual screening in drug
  discovery: a prospective quantum advantage.
\newblock {\em Machine Learning: Science and Technology}, 4(1):015023, feb
  2023.

\bibitem{PhysRevResearch.3.033221}
Sau~Lan Wu, Shaojun Sun, Wen Guan, Chen Zhou, Jay Chan, Chi~Lung Cheng, Tuan
  Pham, Yan Qian, Alex~Zeng Wang, Rui Zhang, Miron Livny, Jennifer Glick,
  Panagiotis~Kl. Barkoutsos, Stefan Woerner, Ivano Tavernelli, Federico
  Carminati, Alberto Di~Meglio, Andy C.~Y. Li, Joseph Lykken, Panagiotis
  Spentzouris, Samuel Yen-Chi Chen, Shinjae Yoo, and Tzu-Chieh Wei.
\newblock Application of quantum machine learning using the quantum kernel
  algorithm on high energy physics analysis at the lhc.
\newblock {\em Phys. Rev. Res.}, 3:033221, Sep 2021.

\bibitem{Li_2023}
Teng Li, Zhipeng Yao, Xingtao Huang, Jiaheng Zou, Tao Lin, and Weidong Li.
\newblock Application of the quantum kernel algorithm on the particle
  identification at the besiii experiment.
\newblock {\em Journal of Physics: Conference Series}, 2438(1):012071, feb
  2023.

\bibitem{9643469}
Marco Pistoia, Syed~Farhan Ahmad, Akshay Ajagekar, Alexander Buts, Shouvanik
  Chakrabarti, Dylan Herman, Shaohan Hu, Andrew Jena, Pierre Minssen, Pradeep
  Niroula, Arthur Rattew, Yue Sun, and Romina Yalovetzky.
\newblock Quantum machine learning for finance iccad special session paper.
\newblock In {\em 2021 IEEE/ACM International Conference On Computer Aided
  Design (ICCAD)}, pages 1--9, 2021.

\bibitem{9915517}
Michele Grossi, Noelle Ibrahim, Voica Radescu, Robert Loredo, Kirsten Voigt,
  Constantin von Altrock, and Andreas Rudnik.
\newblock Mixed quantum–classical method for fraud detection with quantum
  feature selection.
\newblock {\em IEEE Transactions on Quantum Engineering}, 3:1--12, 2022.

\bibitem{suzuki_analysis_2020}
Yudai Suzuki, Hiroshi Yano, Qi~Gao, Shumpei Uno, Tomoki Tanaka, Manato Akiyama,
  and Naoki Yamamoto.
\newblock Analysis and synthesis of feature map for kernel-based quantum
  classifier.
\newblock {\em Quantum Machine Intelligence}, 2(1):9, July 2020.

\bibitem{PhysRevA.102.032420}
Ryan LaRose and Brian Coyle.
\newblock Robust data encodings for quantum classifiers.
\newblock {\em Phys. Rev. A}, 102:032420, Sep 2020.

\bibitem{PhysRevA.101.032308}
Maria Schuld, Alex Bocharov, Krysta~M. Svore, and Nathan Wiebe.
\newblock Circuit-centric quantum classifiers.
\newblock {\em Phys. Rev. A}, 101:032308, Mar 2020.

\bibitem{9259210}
T.~M.~L. {Veras}, I.~C.~S. {De Araujo}, K.~D. {Park}, and A.~J. {da Silva}.
\newblock Circuit-based quantum random access memory for classical data with
  continuous amplitudes.
\newblock {\em IEEE Transactions on Computers}, pages 1--1, 2020.

\bibitem{araujo_divide-and-conquer_2021}
Israel~F. Araujo, Daniel~K. Park, Francesco Petruccione, and Adenilton~J.
  da~Silva.
\newblock A divide-and-conquer algorithm for quantum state preparation.
\newblock {\em Scientific Reports}, 11(1):6329, March 2021.

\bibitem{araujo_configurable_2023}
Israel~F. Araujo, Daniel~K. Park, Teresa~B. Ludermir, Wilson~R. Oliveira,
  Francesco Petruccione, and Adenilton~J. Da~Silva.
\newblock Configurable sublinear circuits for quantum state preparation.
\newblock {\em Quantum Information Processing}, 22(2):123, 2023.

\bibitem{pesah2020absence}
Arthur Pesah, M.~Cerezo, Samson Wang, Tyler Volkoff, Andrew~T. Sornborger, and
  Patrick~J. Coles.
\newblock Absence of barren plateaus in quantum convolutional neural networks.
\newblock {\em Phys. Rev. X}, 11:041011, Oct 2021.

\bibitem{hur2022quantum}
Tak Hur, Leeseok Kim, and Daniel~K Park.
\newblock Quantum convolutional neural network for classical data
  classification.
\newblock {\em Quantum Machine Intelligence}, 4(1):3, 2022.

\bibitem{kim2023classical}
Juhyeon Kim, Joonsuk Huh, and Daniel~K. Park.
\newblock Classical-to-quantum convolutional neural network transfer learning.
\newblock {\em Neurocomputing}, 555:126643, 2023.

\bibitem{oh2023quantum}
Hyeondo Oh and Daniel~K Park.
\newblock Quantum support vector data description for anomaly detection.
\newblock {\em arXiv preprint arXiv:2310.06375}, 2023.

\bibitem{lecun2010mnist}
Yann LeCun, Corinna Cortes, and CJ~Burges.
\newblock Mnist handwritten digit database.
\newblock {\em ATT Labs [Online]. Available: http://yann.lecun.com/exdb/mnist},
  2, 2010.

\bibitem{jerbi_quantum_2023}
Sofiene Jerbi, Lukas~J. Fiderer, Hendrik Poulsen~Nautrup, Jonas~M. Kübler,
  Hans~J. Briegel, and Vedran Dunjko.
\newblock Quantum machine learning beyond kernel methods.
\newblock {\em Nature Communications}, 14(1):517, January 2023.

\bibitem{xiao2017/online}
Han Xiao, Kashif Rasul, and Roland Vollgraf.
\newblock Fashion-mnist: a novel image dataset for benchmarking machine
  learning algorithms, 2017.

\bibitem{berezniuk2020scale}
Oksana Berezniuk, Alessio Figalli, Raffaele Ghigliazza, and Kharen Musaelian.
\newblock A scale-dependent notion of effective dimension.
\newblock {\em arXiv preprint arXiv:2001.10872}, 2020.

\bibitem{abbas2021_b}
Amira Abbas, David Sutter, Alessio Figalli, and Stefan Woerner.
\newblock {Effective dimension of machine learning models}.
\newblock {\em arXiv}, 2021.

\bibitem{holmes_2022}
Zoë Holmes, Kunal Sharma, M.~Cerezo, and Patrick~J. Coles.
\newblock {Connecting Ansatz Expressibility to Gradient Magnitudes and Barren
  Plateaus}.
\newblock {\em PRX Quantum}, 3(1):010313, 2022.

\bibitem{PhysRevLett.117.080501}
Michael~J. Bremner, Ashley Montanaro, and Dan~J. Shepherd.
\newblock Average-case complexity versus approximate simulation of commuting
  quantum computations.
\newblock {\em Phys. Rev. Lett.}, 117:080501, Aug 2016.

\bibitem{bergholm2020pennylane}
Ville Bergholm, Josh Izaac, Maria Schuld, Christian Gogolin, M.~Sohaib Alam,
  Shahnawaz Ahmed, Juan~Miguel Arrazola, Carsten Blank, Alain Delgado, Soran
  Jahangiri, Keri McKiernan, Johannes~Jakob Meyer, Zeyue Niu, Antal Száva, and
  Nathan Killoran.
\newblock Pennylane: Automatic differentiation of hybrid quantum-classical
  computations.
\newblock {\em arXiv preprint arXiv:1811.04968}, 2020.

\bibitem{Sim_expressibility}
Sukin Sim, Peter~D. Johnson, and Alán Aspuru-Guzik.
\newblock Expressibility and entangling capability of parameterized quantum
  circuits for hybrid quantum-classical algorithms.
\newblock {\em Advanced Quantum Technologies}, 2(12):1900070, 2019.

\bibitem{broughton2020tensorflow}
Michael Broughton, Guillaume Verdon, Trevor McCourt, Antonio~J Martinez,
  Jae~Hyeon Yoo, Sergei~V Isakov, Philip Massey, Ramin Halavati, Murphy~Yuezhen
  Niu, Alexander Zlokapa, et~al.
\newblock Tensorflow quantum: A software framework for quantum machine
  learning.
\newblock {\em arXiv preprint arXiv:2003.02989}, 2020.

\bibitem{Mari2020transferlearningin}
Andrea Mari, Thomas~R. Bromley, Josh Izaac, Maria Schuld, and Nathan Killoran.
\newblock Transfer learning in hybrid classical-quantum neural networks.
\newblock {\em {Quantum}}, 4:340, October 2020.

\bibitem{co-design}
Weiwen Jiang, Jinjun Xiong, and Yiyu Shi.
\newblock A co-design framework of neural networks and quantum circuits towards
  quantum advantage.
\newblock {\em Nature Communications}, 12(1):579, 2021.

\bibitem{9951229}
Minzhao Liu, Junyu Liu, Rui Liu, Henry Makhanov, Danylo Lykov, Anuj Apte, and
  Yuri Alexeev.
\newblock Embedding learning in hybrid quantum-classical neural networks.
\newblock In {\em 2022 IEEE International Conference on Quantum Computing and
  Engineering (QCE)}, pages 79--86, 2022.

\bibitem{larocca2022group}
Mart{\'\i}n Larocca, Fr{\'e}d{\'e}ric Sauvage, Faris~M Sbahi, Guillaume Verdon,
  Patrick~J Coles, and Marco Cerezo.
\newblock Group-invariant quantum machine learning.
\newblock {\em PRX Quantum}, 3(3):030341, 2022.

\bibitem{meyer2023exploiting}
Johannes~Jakob Meyer, Marian Mularski, Elies Gil-Fuster, Antonio~Anna Mele,
  Francesco Arzani, Alissa Wilms, and Jens Eisert.
\newblock Exploiting symmetry in variational quantum machine learning.
\newblock {\em PRX Quantum}, 4(1):010328, 2023.

\bibitem{vapnik1999nature}
Vladimir Vapnik.
\newblock {\em The nature of statistical learning theory}.
\newblock Springer science \& business media, 1999.

\bibitem{incudini2022structure}
Massimiliano Incudini, Francesco Martini, and Alessandra Di~Pierro.
\newblock Structure learning of quantum embeddings.
\newblock {\em arXiv preprint arXiv:2209.11144}, 2022.

\bibitem{kwon2023feature}
Hyeokjea Kwon, Hojun Lee, and Joonwoo Bae.
\newblock Feature map for quantum data: Probabilistic manipulation.
\newblock {\em arXiv preprint arXiv:2303.15665}, 2023.

\bibitem{kandala_hardware-efficient_2017}
Abhinav Kandala, Antonio Mezzacapo, Kristan Temme, Maika Takita, Markus Brink,
  Jerry~M. Chow, and Jay~M. Gambetta.
\newblock Hardware-efficient variational quantum eigensolver for small
  molecules and quantum magnets.
\newblock {\em Nature}, 549(7671):242--246, September 2017.

\bibitem{scikit-learn}
F.~Pedregosa, G.~Varoquaux, A.~Gramfort, V.~Michel, B.~Thirion, O.~Grisel,
  M.~Blondel, P.~Prettenhofer, R.~Weiss, V.~Dubourg, J.~Vanderplas, A.~Passos,
  D.~Cournapeau, M.~Brucher, M.~Perrot, and E.~Duchesnay.
\newblock Scikit-learn: Machine learning in {P}ython.
\newblock {\em Journal of Machine Learning Research}, 12:2825--2830, 2011.

\bibitem{kingma2017adam}
Diederik~P Kingma and Jimmy Ba.
\newblock Adam: A method for stochastic optimization.
\newblock {\em arXiv preprint arXiv:1412.6980}, 2017.

\bibitem{qiskit2024}
Ali Javadi-Abhari, Matthew Treinish, Kevin Krsulich, Christopher~J. Wood, Jake
  Lishman, Julien Gacon, Simon Martiel, Paul~D. Nation, Lev~S. Bishop,
  Andrew~W. Cross, Blake~R. Johnson, and Jay~M. Gambetta.
\newblock Quantum computing with {Q}iskit, 2024.

\end{thebibliography}
\end{document}